\documentclass[8.5pt,twoside]{article}

\usepackage[super,sort&compress,comma]{natbib} 
\usepackage{times,mathptm}
\usepackage{sectsty}
\usepackage{balance} 
\usepackage[text={11.3cm,20.4cm},centering]{geometry} 

\usepackage{graphicx} 
\usepackage{lastpage}
\usepackage[format=plain,justification=raggedright,singlelinecheck=false,font=small,labelfont=bf,labelsep=space]{caption} 
\usepackage{fancyhdr}
\begin{document}
\def\gtrsim{\mathrel{\hbox{\rlap{\hbox{\lower4pt\hbox{$\sim$}}}\hbox{$>$}}}}
\def\lesssim{\mathrel{\hbox{\rlap{\hbox{\lower4pt\hbox{$\sim$}}}\hbox{$<$}}}}
\newcommand{\cc}{\mbox{cm$^{-3}$}}

\thispagestyle{plain}
\fancypagestyle{plain}{
\fancyhead[L]{\includegraphics[height=8pt]{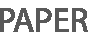}}
\fancyhead[C]{\hspace{-1cm}\includegraphics[height=15pt]{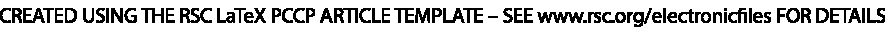}}
\fancyhead[R]{\includegraphics[height=10pt]{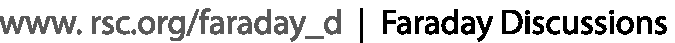}\vspace{-0.2cm}}
\renewcommand{\headrulewidth}{1pt}}
\renewcommand{\thefootnote}{\fnsymbol{footnote}}
\renewcommand\footnoterule{\vspace*{1pt}%
\hrule width 11.3cm height 0.4pt \vspace*{5pt}} 
\setcounter{secnumdepth}{5}

\makeatletter 
\renewcommand{\fnum@figure}{\textbf{Fig.~\thefigure~~}}
\def\subsubsection{\@startsection{subsubsection}{3}{10pt}{-1.25ex plus -1ex minus -.1ex}{0ex plus 0ex}{\normalsize\bf}} 
\def\paragraph{\@startsection{paragraph}{4}{10pt}{-1.25ex plus -1ex minus -.1ex}{0ex plus 0ex}{\normalsize\textit}} 
\renewcommand\@biblabel[1]{#1}            
\renewcommand\@makefntext[1]%
{\noindent\makebox[0pt][r]{\@thefnmark\,}#1}
\makeatother 
\sectionfont{\large}
\subsectionfont{\normalsize} 

\fancyfoot{}
\fancyfoot[LO,RE]{\vspace{-7pt}\includegraphics[height=8pt]{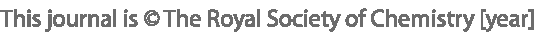}}
\fancyfoot[CO]{\vspace{-7pt}\hspace{5.9cm}\includegraphics[height=7pt]{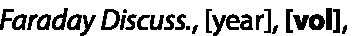}}
\fancyfoot[CE]{\vspace{-6.6pt}\hspace{-7.2cm}\includegraphics[height=7pt]{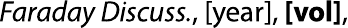}}
\fancyfoot[RO]{\scriptsize{\sffamily{1--\pageref{LastPage} ~\textbar  \hspace{2pt}\thepage}}}
\fancyfoot[LE]{\scriptsize{\sffamily{\thepage~\textbar\hspace{3.3cm} 1--\pageref{LastPage}}}}
\fancyhead{}
\renewcommand{\headrulewidth}{1pt} 
\renewcommand{\footrulewidth}{1pt}
\setlength{\arrayrulewidth}{1pt}
\setlength{\columnsep}{6.5mm}
\setlength\bibsep{1pt}

\noindent\LARGE{\textbf{Exploring the Origins of Carbon in Terrestrial Worlds$^\dag$}}
\vspace{0.6cm}

\noindent\large{\textbf{Edwin Bergin\textit{$^{a}$}, L. Ilsedore Cleeves\textit{$^{a}$}},
\textbf{Nathan Crockett\textit{$^{b}$}, and Geoffrey Blake\textit{$^{b}$}}
\vspace{0.5cm}

\noindent\textit{\small{\textbf{Received 16th January 2014,  Accepted 12th February 2014\newline
First published on the web 12th February 2014}}}

\noindent \textbf{\small{DOI: 10.1039/c4fd00003j}}
\vspace{0.6cm}

\noindent \normalsize{Given the central role of carbon in the chemistry of life, it is a fundamental question as to how carbon is supplied to the Earth, in what form and when.  
We provide an accounting of carbon found in solar system bodies, in particular a comparison between the organic content of meteorites and that in identified organics in the dense interstellar medium (ISM).  Based on this accounting identified organics created by the chemistry of star formation could contain at most $\sim$15\% of the organic carbon content in primitive meteorites and significantly less for cometary organics, which represent the putative contributors to starting materials for the Earth. 
  In the ISM $\sim30\%$
  of the elemental carbon is found in 
  CO, either in the gas or ices,
  with a typical abundance of $\sim10^{-4}$ (relative to H$_2$).  Recent observations of the TW Hya disk find that the gas phase abundance of CO is reduced by an order of magnitude compared to this value.  We explore  a solution where the volatile CO is destroyed via a gas phase processes, providing an additional source of carbon for organic material to be incorporated into planetesimals and cometesimals.   This chemical processing mechanism requires warm grains ($>$ 20~K), partially ionized gas, and sufficiently small ($a_{\rm grain}<10\; \mu$m) grains, i.e. a larger {\em total} grain surface area, such that freeze-out is efficient. 
 Under these conditions static (non-turbulent) 
 chemical models predict that a large fraction of the carbon nominally sequestered in CO can be the source of carbon for a wide variety of organics that are present as ice coatings on the surfaces of warm pre-planetesimal dust grains.
  }
\vspace{0.5cm}

\section{Introduction}


\footnotetext{\textit{$^{a}$~University of Michigan, Department of Astronomy, 500 Church St., Ann Arbor, MI 48109 USA. Fax:1 734 763 6317  Tel:1 734 615 8720, E-mail: ebergin@umich.edu;
\textit{$^{b}$California Institute of Technology, Division of Geological \& Planetary
Sciences, MS 150-21, Pasadena, CA 91125, USA}}}



The birth of terrestrial worlds involves the accumulation of material over a potentially wide range of distances in the young pre-planetary solar nebula disk (e.g. \citet[][]{Morby12}. and references therein)  Our general expectation is thus that Earth formed from material both interior to and beyond its location at 1 Astronomical Unit from the Sun.    This is important as prevailing theories argue that, due to the hot temperatures of pre-planetary materials,  the Earth formed dry with little water.  
For that reason, the Earth is thought to have received water from hydrated rocks (asteroids) and perhaps icy planetesimals that exist beyond the so-called solar nebula disk\footnote{In this contribution we will refer to the disk of material out of which our planetary system as formed as the solar nebula disk and other extra-solar disk systems as protoplanetary disks.} snow line; although alternative theories exist (see summary presented by \citet[][]{evd_ppvi}).

We aim to explore some of the chemical/physical mechanisms associated with a potentially related topic:  how the Earth received its carbon.  Here we will draw upon previous works\citep{lbn10, pont_ppvi} to discuss the ``carbon-problem'' that outlines the likelihood that the Earth (using meteoritic material as a proxy) is carbon-poor relative to carbon available at birth.    Two additional key facets are notable:

\begin{enumerate}

\item Most carbon in the solar nebular disk was in volatile form.   In the interstellar medium carbon monoxide accounts for $\sim30$\% of elemental carbon.  Thus it is reasonable to assume that a large fraction of the volatile carbon in a disk resides in gaseous or solid state CO.

\item Most meteoritic and much of cometary carbon is organic in form as opposed to volatile CO, CO$_2$, or even CH$_3$OH.  In this regard, meteorites represent a likely source of carbon, and other volatiles such as water, to the young Earth\citep{Morby12, Marty12, Albarede13}.    
\end{enumerate}

In this contribution we explore potential connections between volatile CO and organic/molecular ices that might contribute to pre-planetary rocks.   This exploration is motivated by a recent indirect measurement of the CO abundance in a nearby disk\citep{favre13a} showing that CO is not a significant reservoir ($<10\%$) of elemental carbon in layers where CO is expected to be in the gas (i.e. not a result of CO freeze-out). 
Thus carbon is somehow processed from volatile CO and potentially into the form of pre-meteoritic organics.     \citet{favre13a} hypothesized that one mechanism for this to occur is via the X-ray ionization of He in the surface layers of the disk.  The resulting He$^+$ atoms can react with gaseous CO and gradually extract the carbon, which can then be processed via disk chemistry 
into more complex, and less volatile, forms that can locally freeze onto the surface of cold dust grains.  This process was first noted in an earlier Faraday discussion by \citet{Aikawa98}. Here we present new models that directly illustrate this effect and its key dependencies on sources of disk ionization and the surface area of small grains.   

In \S 2 we will outline the carbon problem in the context of the composition of bodies in the solar system and interstellar medium.  We also explore the question of how much of the organic carbon in solar system bodies might originate in the interstellar medium.  We will also summarize and expand upon the \citet{favre13a} result to motivate the inference that chemical processing of volatile CO appears to be active. 
in \S 3 we will present new results from a chemical model that explores the time dependence of CO chemical processing in the context of a realistic disk model and in \S 4 we will summarize the implications for the overall chemical evolution of carbon within pre-planetary materials.

\section{Carbon in the Solar System, ISM, and Protoplanetary Disks}
\label{sec:carbon}

\subsection{Carbon in the Solar System and the Interstellar Medium}

Fig.~\ref{fig:c} illustrates the dramatic differences in the amount of elemental carbon relative to silicon incorporated into planets, comets, and asteroids represented by meteorites.     Exploring the relative amounts of carbon, we see that icy bodies (i.e. comets) contain an abundance of carbon in the form of ices and dust (perhaps amorphous carbon) comparable to that seen in the Sun. 
In contrast, the Earth's mantle has many orders of magnitude less carbon (relative to Si) than was available during formation as represented by the Sun or the diffuse ISM.  
The  Earth's mantle is predominantly comprised of silicate minerals with a total carbon content of $\sim 10^{23}$ g \citep{dh10}.
This is a tiny fraction of the Earth's mass ($\sim 0.002$\%), but we lack knowledge regarding the carbon content in the Earth's core.  Thus while the mantle is clearly carbon-poor relative to the elemental carbon available at birth, we cannot make definitive statements regarding the bulk Earth.    Meteorites, in particular undifferentiated carbonaceous chondrites (class CI), have long been posited as tracers of the starting materials of the Earth and these rocks have an order of magnitude depletion of carbon relative to the amount available at formation. 

\begin{figure}[h]
  \includegraphics[height=8cm]{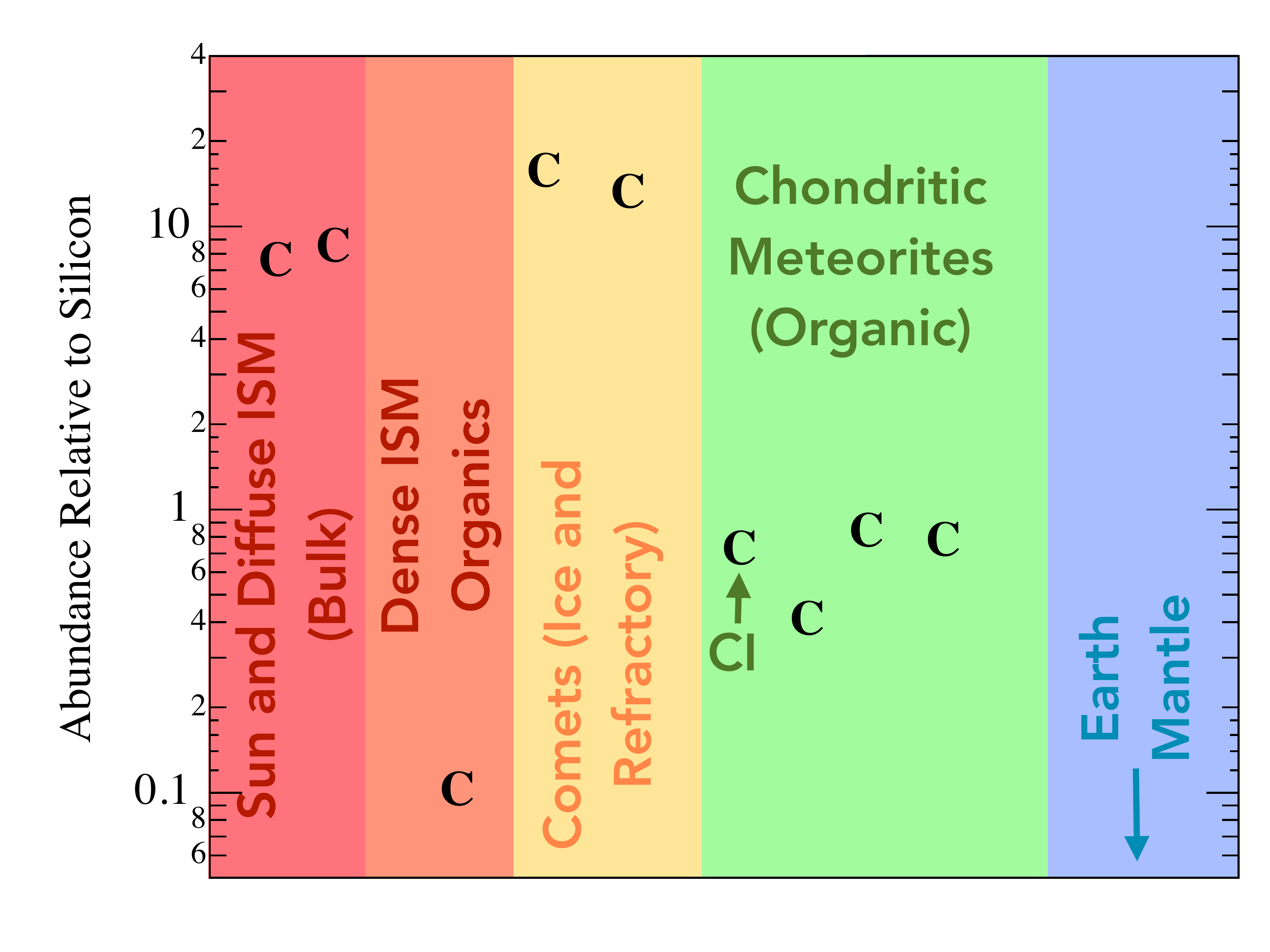}
  \caption{Plot of total elemental carbon abundance in various solar system bodies
  (Sun; comets Hale-Bopp and Halley; 4 chondritic meteorites, classes CI, CV, CM, CO; Earth mantle);  
  and the interstellar medium relative to elemental silicon.   Figure adapted from \citet{lbn10} and \citet{pont_ppvi}
  and references for abundance estimates are given in those publications.   The abundances of dense
  ISM organics is taken from \citet{Crockett14a} and compiled for the first time here.
}
  \label{fig:c}
\end{figure}

The severe depletions of C suggest that carbon must have been present in fairly volatile forms in the solar nebula disk.   One suggestion is that Earth's carbon was supplied from meteoritic organics\citep{Marty12, Albarede13}. 
We present a new perspective on this issue through the analysis of the most complete spectral observations (in terms of spectral coverage) of the star-forming molecular gas inventory in the Herschel spectrum of Orion KL \citep{Crockett14a}.  
 This region is the nearest example of massive star formation (an environment analogous to that in which the Sun was thought to have formed)\citep{solarbirth}. 
 The unique Orion spectrum is the first example in which we can examine a near complete chemical census (to observational sensitivity limits) in the highest temperature gas in which we know the ices have been fully evaporated. 
Since organics are theorized to form on the surfaces of dust as ices, we are therefore tracing the composition of simple organics provided (as ice) to forming stellar and planetary systems.
 
For this comparison we have summed up the carbon abundances in detected organics to provide a total organic carbon abundance relative to hydrogen and normalized these to the silicon abundance from present day cosmic abundance estimates \citep{Nieva12}.  
Orion KL abundances were calculated from the column densities estimated by matching emission from numerous transitions and normalization to the total gas (H$_2$) column using over 7 transitions of C$^{18}$O and also C$^{17}$O (taken from \citet{Plume12}); thus these abundances are very robust.  The errors are dominated by the beam couplings between the detected organic and the Herschel beam (which we derived from the ALMA science verification data of Orion) and are likely below $\sim$25\%.   For this comparison we have summed up the carbon abundances in detected organics to provide a total organic carbon abundance relative to Hydrogen and normalized these to the Silicon abundance from present day cosmic abundance estimates \citep{Nieva12}, assuming all Silicon resides in dust grains.


Exploring the most abundant, detected, and identified organics in space we find that the chemistry associated with stellar birth might have created up to  $\sim$15\% of the organics that could be placed into meteorites.   
Carbon in comets is found primarily in the dust with a division of 2:1 between dust:ice\citep{Fomenkova99}, at least in comet Halley, which has {\em in situ} measurements.  Furthermore, about 50\% 
of the refractory carbon
is found in complex organic form\citep{Fomenkova99}, which means that the amount of carbon contained in organic (refractory) form is $\gtrsim 5$ times that of CI chondrites and significantly higher than the identified ISM organic molecules. Now this accounting is based solely on identified organic molecules and one star-forming region.  However, it is also the most complete accounting of the extent of interstellar chemistry to date and it suggests that the chemistry occurring prior to stellar birth both in the gas and ice cannot account for organic material in the solar system record. We are therefore ``missing'' carbon, which is a significant astrobiological question, if one considers that astronomically we wish to characterize the starting materials. Instead the carbon found in solar system rocks resides in PAHs to some type of carbonaceous grains, which may originate primarily in the interstellar medium. 
However, there might also be subsequent processing in the disk environment which could rearrange major carbon carriers and in particular extract carbon from volatiles and place it into different forms, which will be addressed in the following section.

\subsection{Volatile Carbon Abundance in the TW Hya Protoplanetary Disk}

In general, the determination of chemical abundances in protoplanetary disks is inherently uncertain.  This is mainly due to the difficulty of measuring the molecular hydrogen content of the the disk, i.e. the total gas mass.    Because H$_2$ does not emit at the temperatures characteristic of much of the disk we are forced to use calibrated proxies, which often have large uncertainties;  see discussion in \citet[][]{bergin_hd} and \citet{williams_araa}.
For TW Hya, the closest planet-forming disk at 51 pc\citep{Mamajek05}, the determination of chemical abundances is more directly enabled via the detection of the fundamental transition of HD at $\sim112\;\mu$m by \citet{bergin_hd}.
 In the case of HD, the HD/H$_2$ ratio is well characterized to be $3.0\pm0.2\times10^{-5}$ by \citet{linsky98}.
With rotational energy spacings better matched to the gas temperature and a weak dipole,  the emission of HD J = 1--0  is in  local thermodynamic equilibrium (LTE) and can readily be converted to a gas mass.  However, while HD co-exists with H$_2$, the HD J = 1--0 transition has a $\Delta E_{1-0} = 128.5$~K, and as a direct result, this transition does not emit appreciably at temperatures below 20~K.   The total estimation of the H$_2$ disk-mass thus requires knowledge of the gas thermal structure\citep{gorti11, bergin_hd}.   However, for the determination of the volatile CO abundance this weakness becomes a strength as gaseous CO is believed to freeze onto grains at temperatures below $\sim$20~K, and thus the emission from both species directly traces the same warm gas. 


Thus \citet{favre13a} used the ratio of optically thin C$^{18}$O emission to that of HD in TW Hya (with isotopic ratios) to provide a measure of the CO abundance  in the so-called warm molecular layer \citep{aikawa_vanz02}.
The estimated CO abundance is $\lesssim10^{-5}$, an order of magnitude below the expected value in the ISM; a level that has been confirmed by an independent analysis\citep{wb13}.    \citet{favre13a} discussed a variety of potential pitfalls and theories to explain this low abundance. We explore two of these here: differences in the self-shielding between HD and C$^{18}$O and chemical processing.

In the former case HD and C$^{18}$O are both capable of self-shielding from the destructive effects of ultraviolet photons (including the effects of mutual shielding via H$_2$ and CO).   If HD reaches full shielding at a layer significantly above that of C$^{18}$O then the abundance measurement would be a lower limit as HD would trace more mass than C$^{18}$O.    To explore this question in Fig.~\ref{fig:ss} we show the self-shielding factor $f$ for each molecule, as a function of the total H$_2$ column, using the self-shielding expressions provided by \citet{Wolcott-green11} for HD and \citet{visser09} for C$^{18}$O, and a disk vertical density profile at $R=20$~AU (see model details below).   This demonstrates that HD does indeed self-shield prior to C$^{18}$O (at N(H$_2$) $\sim10^{21}$~cm$^{-2}$ and $\sim10^{22}$~cm$^{-2}$, respectively).   However, while this difference in self-shielding column exists, the total column density of gas at $R=20$~AU is N(H$_2$) $=7\times10^{24}$~cm$^{-2}$ and is above 20~K at all heights.  Thus the mass contained in this layer where the differential self shielding occurs constitutes a negligible fraction of the mass ($<1\%$) and cannot account for the order of magnitude deficit in CO abundance relative to interstellar.

\begin{figure}
\centering
    \includegraphics[width=0.6\textwidth, angle=90]{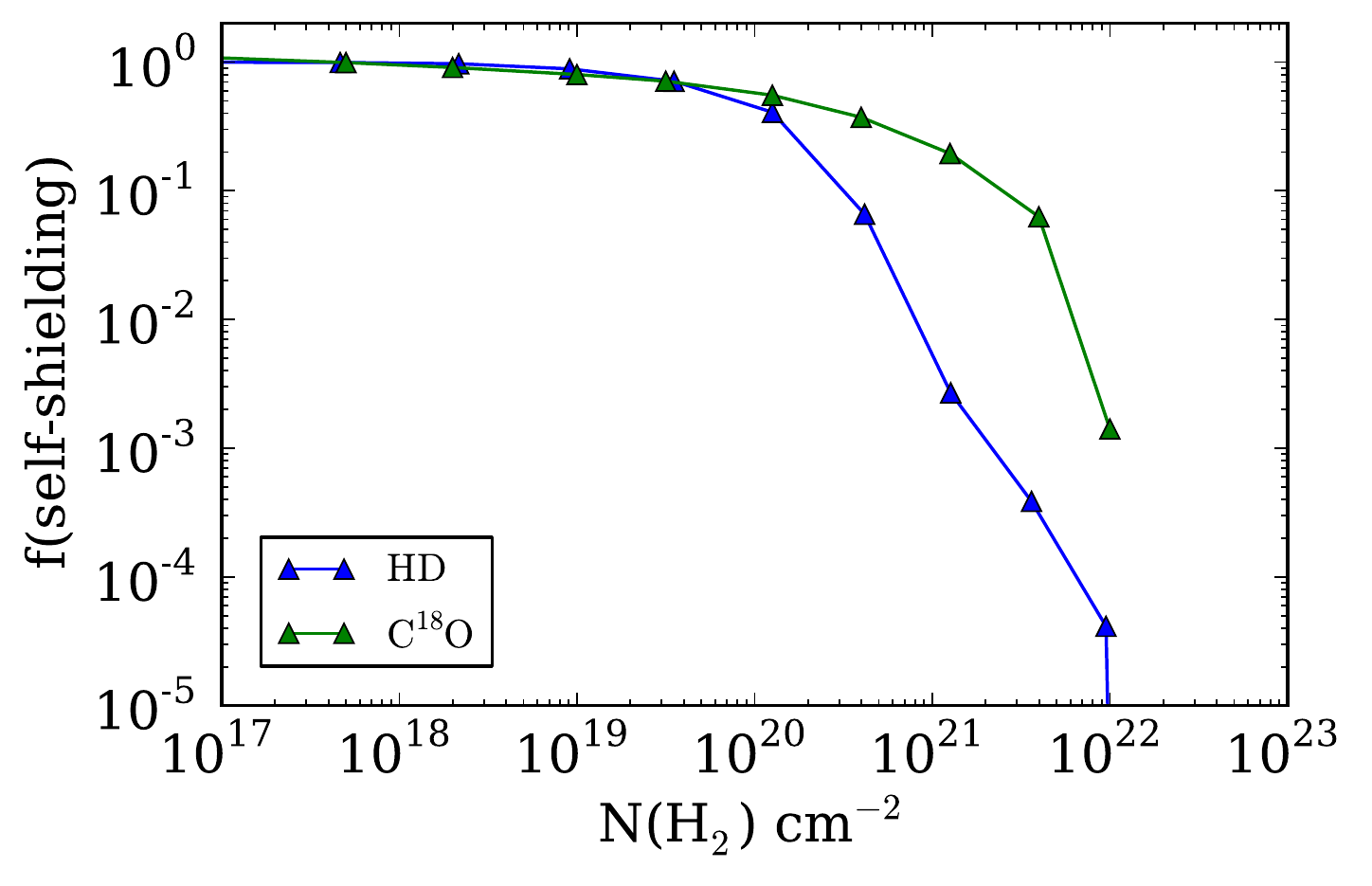}
  \caption{\footnotesize Self-shielding factors for HD and C$^{18}$O as a function of the molecular hydrogen column density.  Density structure taken from a cut in the disk physical model, presented in \S 3.2, at 20 AU.}
  \label{fig:ss}
\end{figure}

The age of TW Hya has some uncertainty, but is believed to be in the range of 3-10 Myr \citep{hoff98, vs11}, which is older than typical T Tauri disk systems\citep{williams_araa}.   Within this timeframe it is possible that there might be disk chemical processing that extracts carbon from volatile CO and places it into molecular material of lower volatility that may contribute to organics incorporated into cometesimals and planetesimals.  In the following sections we will outline a potential theoretical mechanism that will operate in any disk system that is exposed to a source of radiation capable of ionizing helium, in this instance, either X-rays or cosmic rays.

\section{Evolution of the Volatile Carbon Reservoir}

\subsection{Potential CO Reprocessing Mechanism}

A simple mechanism to extract carbon from CO relies on the fact that in dense ($n>10^5$ \cc ) and cold ($T<50$~K) gas the timescales for molecules to collide with dust grains are short, t $< 10^5$ yrs \citep{evd_ppiii, bt_araa, hkbm09}.   In this case the chemistry becomes dominated by gas phase freeze-out as molecules are adsorbed on grain surfaces.  This effect is well documented in disk systems.  In the solar nebula disk the presence of volatile CO in solar system comets clearly shows that pre-cometary ices contained highly volatile species\cite{mc11}.  In extra-solar disks the formation of pre-planetesimal ices is inferred due to the low abundances of various molecules measured in disks compared to their interstellar medium abundances\citep{dgg97, kastner_twhya, hoger11a} and in some cases via direct observations\citep{pont05, Honda09}.

The disk surface is directly exposed to stellar UV and X-ray radiation producing sharp radial and vertical thermal gradients\citep{Kamp04, nomura07}.   This leads to vertical chemical stratification into roughly three domains: (1) the photon-dominated surface where molecules are photodissociated, (2) a ``warm molecular layer'' where dust temperatures are warm enough ($T_d > 20$~K) for CO and other species with similar volatility to be present in UV-shielded gas, (3) the dense ($n \gg 10^5$ \cc ) cold ($T \le 20$~K) midplane where molecules are frozen as ices\citep{aikawa_vanz02, bergin_ppv}.  The ``warm molecular layer''\citep{aikawa_vanz02} should be distinguished from the snow line\citep{hayashi_mmsn}, which typically refers to the radial abundance gradient in the midplane with the snow line located at the evaporation front.

For CO the warm molecular layer exists when the dust temperature is $> 20$~K; however there are a host of more complex carbon-bearing molecules that remain frozen on grain surfaces \citep{collings_cobind, collings_lab}. In this layer, if CO$_2$ remains frozen, the elemental C/O ratio will be close to unity because water remains as ice for temperatures below $\sim 100-150$~K (depending upon pressure)\citep{fraser_h2obind} and the carbon and oxygen resides primarily in CO. If there is a source of radiation capable of ionizing helium atoms then  the following sequence of reactions will extract small amounts of carbon in ionized atomic form: 

\begin{equation}
\rm{He \;+\; \zeta \rightarrow He^+ \;+\; e}
\end{equation}

\begin{equation}
\rm{He^+\; +\; CO \rightarrow C^+ \;+\; O\; +\; He}.
\end{equation}

\noindent  Much of this free carbon reforms as CO in the gas, but a fraction can follow a variety of gas-phase or surface pathways to make less volatile ices such as CO$_2$, hydrocarbons, or CH$_3$OH.   In this fashion the grains operate as a sink to the carbon chemistry as they trap any volatile that has its sublimation temperature below the temperature of dust grains.   Over time the CO abundance might become eroded, a facet noted and discussed by \citet{aikawa97} and elucidated further below. 

\subsection{Chemical Model}
\label{sec:model}

To explore this theory and its dependencies more directly we have used the chemical models of 
\citet{fogel11} and \citet{cleeves11}.   Briefly these models adopt realistic disk physical structures (gas density and temperature, dust temperature, dust properties) motivated by resolved disk observations and the overall spectral energy distribution\citep{williams_araa}.   We adopt the observed UV field of TW Hya\citep{herczeg_twhya1} and X-ray\citep{Preibisch05} radiation fields for T Tauri systems and use a Monte-Carlo radiation transfer code to solve for the position and wavelength dependent UV and X-ray radiation field within the disk.  
The model X-ray flux has a total luminosity of $L_{\rm XR} = 10^{29.5}$ erg s$^{-1}$ between 0.1 -- 10 keV with absorption/scattering cross-sections taken from \citet{bb11b}.
 This code directly includes the propagation of Ly~$\alpha$ photons via H-atom resonant scattering and dust absorption/scattering\citep{bb11a}; Ly~$\alpha$ radiation dominates the stellar UV emission generated by magnetospheric accretion\citep{herczeg_twhya1, Schindhelm12}.   More directly, the adopted physical model is the same as used by \citet{cleeves13a} and the reader is referred to that publication for greater detail.

With the physical structure defined we then use a detailed chemical network\citep{osu04} including gas grain interactions (freeze-out, sublimation, cosmic ray induced desorption, photodesorption) to solve for the time dependence of predicted chemical abundances.    A limited amount of grain surface reactions were included that create H$_2$, H$_2$O, OH, CH$_4$, CO$_2$, H$_2$CO, and CH$_3$OH ice.   Thus we have (roughly) included pathways for the hydrogenation of O, C, and CO on grain surfaces.   This surface chemistry is quite limited compared to the wide variety of potential pathways\citep{hhl92, cr09}, but allows for an exploration of what might happen to the carbon chemistry.    We note that in this regard the end point of the carbon surface chemistry is CH$_3$OH and its ice abundance is thus an upper limit. The detailed physical structure is given in Fig.~\ref{fig:phys}.

\begin{figure}
    \includegraphics[width=0.95\textwidth, angle=0]{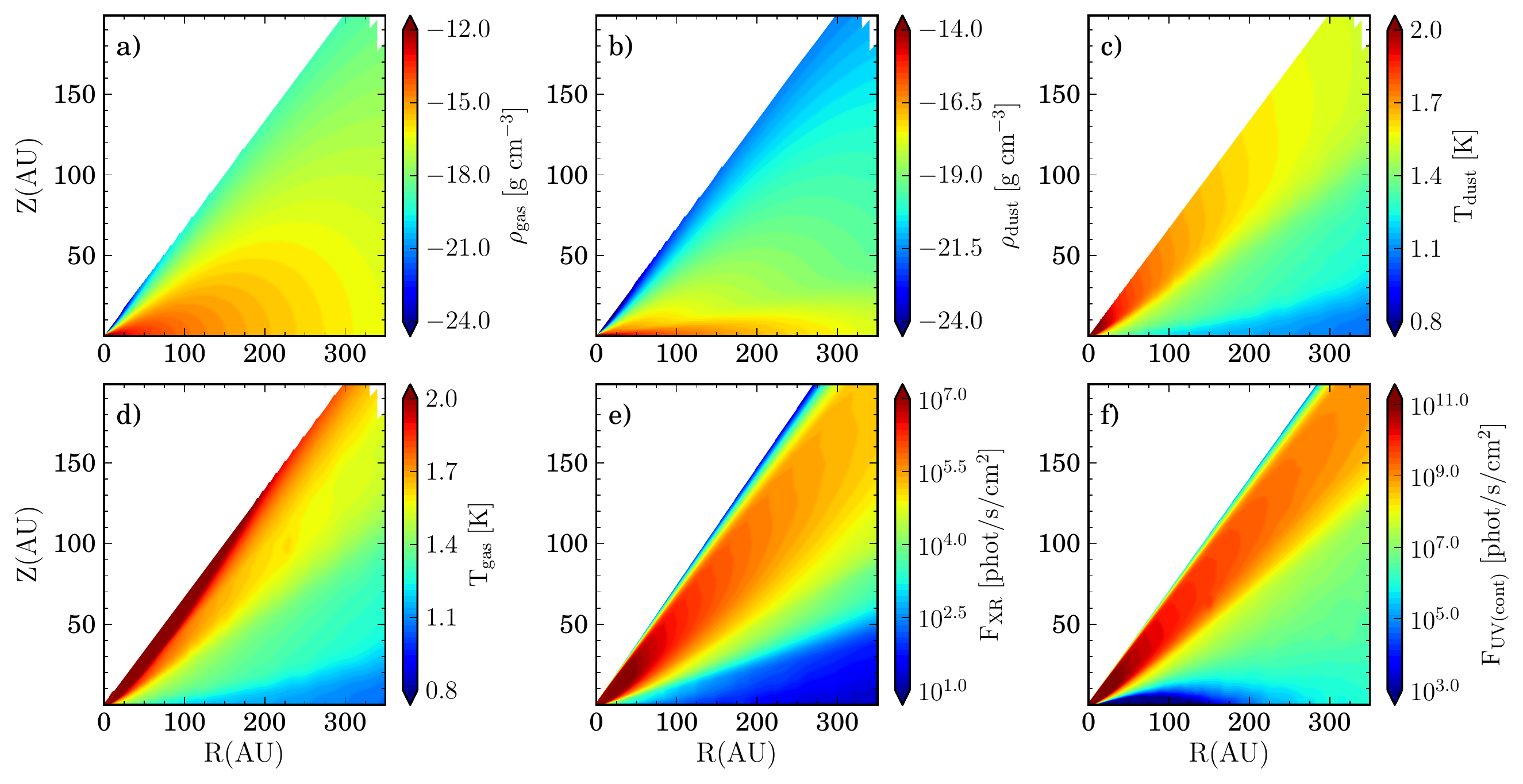}
  \caption{\footnotesize Physical parameters of the adopted disk model (and Figure) taken from \citet{cleeves13a}.   In panel (a) we provide the gas density, (b) the dust density, (c) dust temperature, (d) gas temperature.
  All parameters are shown as a function of disk radial and vertical position.  In panels (e) and (f) we show the derived UV and X-ray photon flux from our simulations.  
 }
  \label{fig:phys}
\end{figure}

The ionization energy of He is 24.59 eV, which requires either X-ray or cosmic ray ionization. 
X-ray ionization dominates the deep disk surface below the UV photon penetration depth and also significantly contributes
to the ionization of the midplane in the inner tens of AU.  Cosmic rays are required to ionize the midplane in the outer (R $>50$ AU) disk\citep{ig99, cleeves13a}. 
  \citet{cleeves13a} explored the question of cosmic ray attenuation in the T Tauri system as cosmic ray penetration can be limited by the stellar magnetic field, stellar winds, and disk winds.  They predict that the cosmic ray ionization rate within disk systems is likely severely reduced compared to the rate in the general interstellar medium.   In this light, since we are using the same model with identical radiation fields, we have adopt their ionization structure and explored two models assuming the disk is exposed to the interstellar cosmic ray radiation field and an attenuated model commensurate with cosmic ray modulation comparable to the Sun at solar maximum.  An additional factor is the timescale for molecular freeze-out onto grain surfaces as the gas-phase processing of CO only occurs when the grains acts as sinks to molecules created in the gas.  The freeze-out timescale is $\tau_{fo} = n_{gr} \sigma_{gr} v$, with $n_{gr}$ the space density of grains, $\sigma_{gr}$ the cross-section, and $v$ the velocity of the gaseous atoms.  $n_{gr} \sigma_{gr}$ is set by the distribution of grain sizes.  The standard grain size distribution in the interstellar medium follows a power-law of $dn(a)/da \propto a^{-3.5}$, with $a$ as the grain radius \citep{mrn}. This is the starting point for disk simulations. Assuming a lower cutoff of 20~\AA , $n_g \sigma_g \simeq 2 \times 10^{-21}n$ cm$^{-1}$, as listed by \citet{hkbm09}, the freeze-out timescale, $\tau_{\,fo}$, is:
 
\begin{equation}
\tau_{\,fo} = 5 \times 10^3 {\rm yrs} \left(\frac{10^5 {\rm~cm^{-3}}}{n}\right) \left(\frac{20 {\rm~K}}{T}\right)^{\frac{1}{2}}\sqrt{A}
\end{equation}

\noindent where $A$ is the molecule mass in atomic mass units.  In our simulations we will adopt this standard value and a second model that assumes grain growth from 0.1 $\mu$m to 10 $\mu$m which increases the freeze-out timescale by a factor of 1000.

  \begin{figure}
\centering
    \includegraphics[width=1.0\textwidth, angle=0]{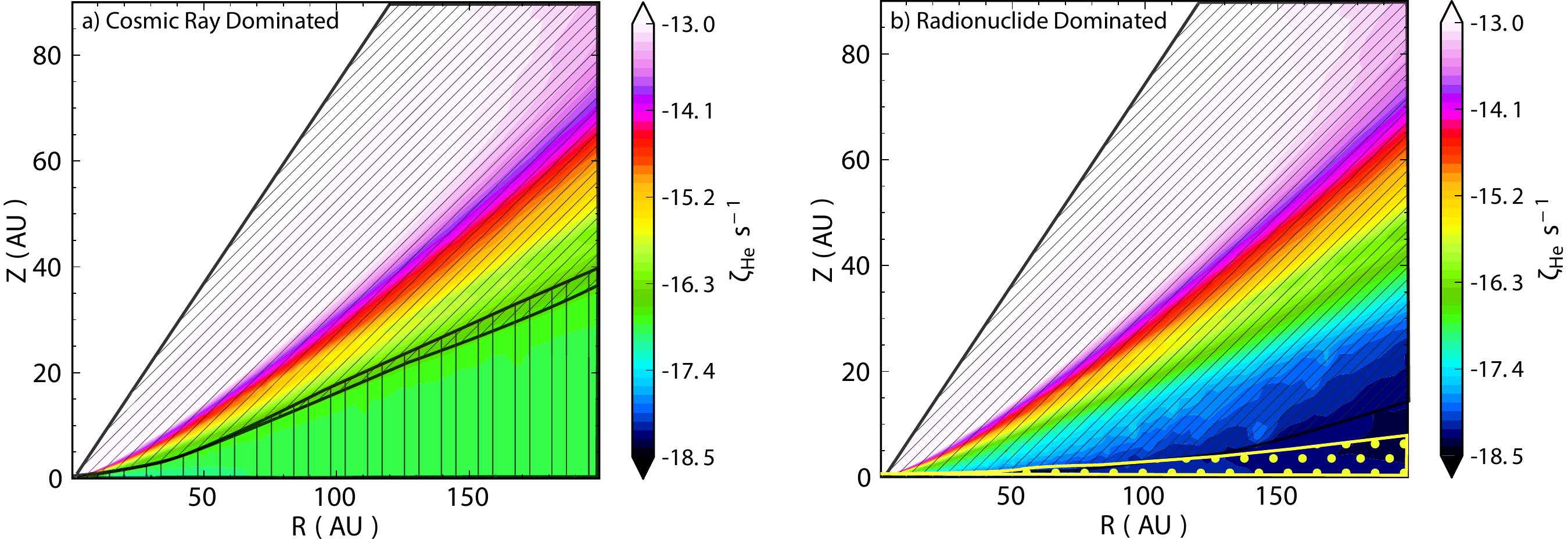}
  \caption{\footnotesize Ionization rate for He atoms as a function of position in the standard disk model.  Two models are shown.  Both models include X-ray ionization and radionuclides as described in \S~\ref{sec:model}.  In panel (a) the model is exposed to the standard interstellar cosmic ray field and in panel (b) the cosmic ray field is reduced as consistent with the attenuation seen in our solar system at the maximum sunspot cycle. In this case, the ``floor'' to the ionization is provided by radionuclides at the initial abundances determined for the solar nebula disk. In each plot diagonal lines denote where the ionization is dominated by X-rays, vertical lines by cosmic rays, and yellow dots dominated by radionuclide decay.}
  \label{fig:ion}
\end{figure}

\subsection{Results}

\subsubsection{He Ionization Rate }

In Fig.~\ref{fig:ion} we show the predicted ionization rate of He atoms using the standard physical model provided in Fig.\ref{fig:phys}.  The two panels show the ionization rate of He assuming the disk is exposed to the standard interstellar cosmic ray radiation field and one that is attenuated due to magnetized stellar winds.  In both cases the disk surface is dominated by X-ray radiation, but the X-ray dominated regime is significantly larger in the case of attenuated cosmic rays.  The major difference lies in the midplane.   In the model with cosmic rays attenuated by a magnetized stellar wind then radionuclides dominate the ionization of the midplane
providing a floor to the ionization rate," is $\sim8\times10^{-19}$~s$^{-1}$ at the midplane at R = 50 AU.
We have assumed solar nebula disk abundances for the radioactive agents, which include $^{26}$Al, $^{36}$Cl and $^{60}$Fe decay\citep{umebayashi09,cleeves13b}.  For a discussion regarding the likelihood that disk systems contain radionuclide ionization see \citet{cleeves13b} and \citet{Jura13}.

A key facet is that as a noble gas, helium is expected to remain in the gas-phase, and, as long as a source of ionization exists, then He$^+$
will be present in the system.  In equilibrium, and in the warm molecular layer where water is frozen as ice, the He$^+$ abundance can be determined by the balance between its ionization rate and destruction with CO and H$_2$.   The neutralizing reaction with H$_2$ has a rate of $k_1 = 7.2 \times 10^{-15}$ cm$^3$ s$^{-1}$ (H$_2^+$ product channel) and $k_2 =  3.7 \times 10^{-14}e^{(-35{\rm~K}/T)}$ cm$^3$ s$^{-1}$  (H$^+$ $+$ H product channel)\citep{Barlow84}, and $k_{\rm CO} = 1.6 \times 10^{-9}$ cm$^3$ s$^{-1}$ is the rate for the reaction between He$^+$ and CO\citep{lhb74}.    In the following we will sum both channels for the reaction between H$^+$ with H$_2$ ($k_{\rm H_2} = k_1 + k_2$) and assume a temperature of 30 K.   Thus in equilibrium, 

\[
n_{\rm He^+} (x_{\rm CO}n_{\rm H_2}k_{\rm CO} + n_{\rm H_2}k_{\rm H_2}) = \zeta_{\rm He} x_{\rm He} n_{\rm H_2}
\]

\noindent where $x_{\rm CO}$ and $x_{\rm He}=0.28$ are the CO and helium abundances respectively, both relative to H$_2$. Solving for $n_{\rm He^+}$ we find,

\[
n_{\rm He^+}  = 0.28\zeta_{\rm He}/(x_{\rm CO}k_{\rm CO} + k_{\rm H_2}).
\]

\noindent The timescale for gas phase CO destruction is then $\tau_{\rm CO} = 1/n_{\rm He^+}k_{\rm CO}$.  A typical ionization rate in the deep surface, where X-ray ionization dominates and the dust temperature is $\sim 30$~K, is $\sim 10^{-15}$ s$^{-1}$ to 10$^{-16}$~s$^{-1}$ (see Fig.~\ref{fig:ion}).   Thus the timescale for CO destruction via reaction with He$^+$ is $\sim 10^{4} - 10^6$ yrs, depending on $\zeta_{\rm He^+}$ and on whether the CO abundance is 10$^{-4}$ (relative to H$_2$) or significantly lower, at which point neutralization with H$_2$ dominates. 

\begin{figure}
\centering
    \includegraphics[width=0.9\textwidth, angle=0]{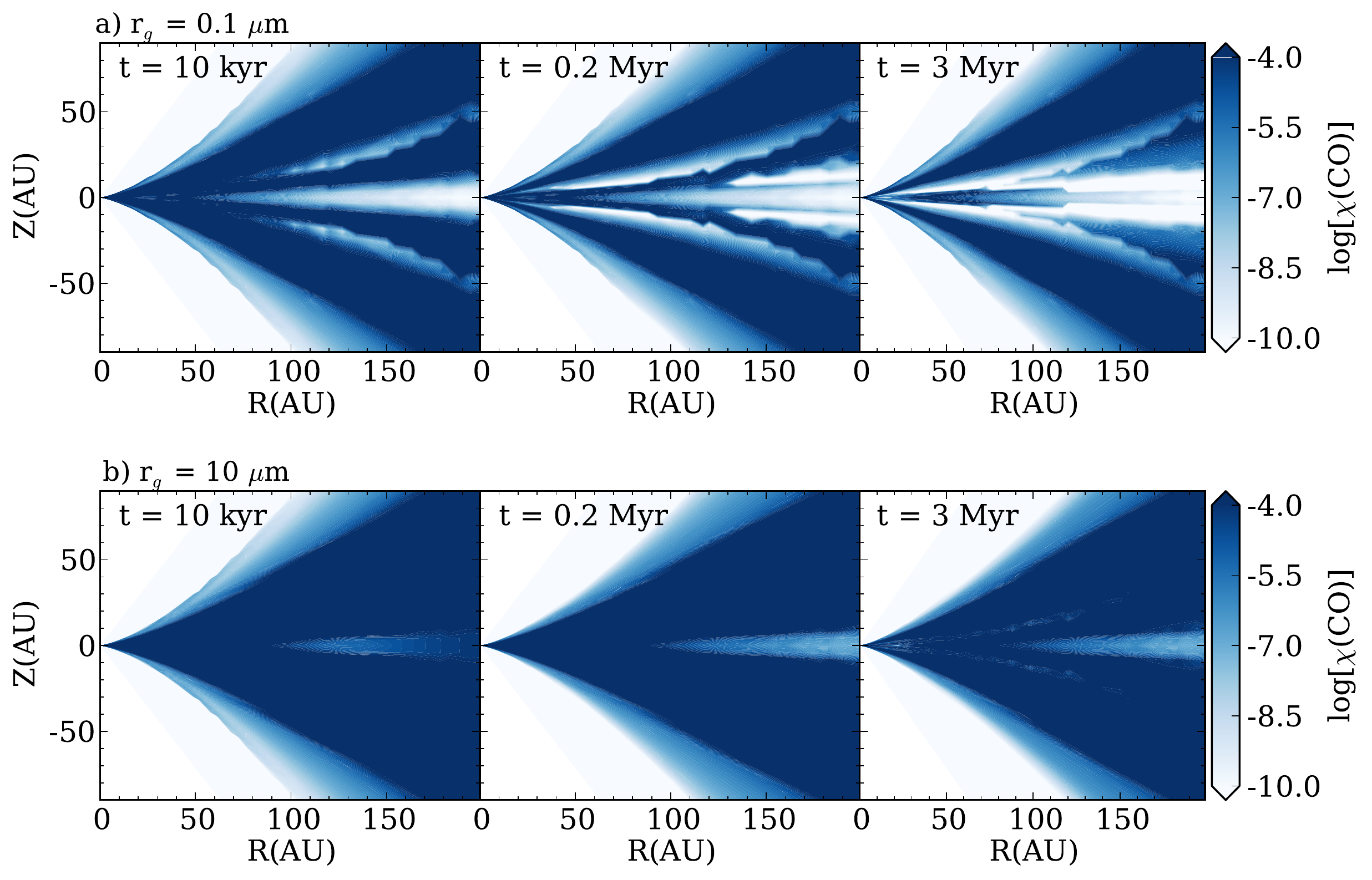}
  \caption{\footnotesize Chemical abundance of CO (relative to H$_2$) as a function of position in the disk
  at selected times.  Top panels are for the standard model with interstellar (0.1$\mu$m) grains and the bottom panels
  assume grain growth to 10$\mu$m size. }
  \label{fig:co}
\end{figure}

\subsubsection{Evolution of CO Abundance }

In Fig.~\ref{fig:co} we provide the time dependence of the CO abundance structure in our standard disk model exposed to an unattenuated interstellar cosmic ray radiation field.   The top panel shows a disk where the grain size distribution is similar to the interstellar medium and the bottom panel a disk that has undergone significant grain evolution.   Focusing first at 10 kyr, the warm molecular layer is seen as the large radial band of ``normal'' CO abundance (10$^{-4}$) on the surface.    In the outer disk (R $>$ 50 AU) significant abundance structure is predicted with two areas of CO depletion.   First is the midplane (Z = 0) where CO is frozen on grains with a snow line, or sublimation front, near 50~AU in this model.   The second zone of CO depletion is the radial band seen in the middle of the warm molecular layer.   This region  is where the reaction with He$^+$ and subsequent gas phase chemical processing is active. 
This exists in the middle of the warm molecular layer and is due to the rapid He$^+$ reaction timescales as the result of X-ray ionization. At later evolutionary steps this zone expands both radially and vertically leading to significant CO depletion in the original warm molecular layer.  Note that the CO gas phase depletion seen near the midplane also expands at later times.  This is not a result of greater freeze-out but rather is gas-phase processing of CO; the lower ionization rates provided by scattered X-rays or cosmic rays lead to longer He$^+$ reaction timescales.

Protoplanetary disks undergo significant grain evolution\citep{natta_ppv, williams_araa}.   
Thus the bottom panels illustrate how this mechanism depends on the evolution of grain surface area with grain growth. 
For larger grain sizes (10 $\mu$m) the chemical processing mechanism becomes inefficient in destroying CO with only minor differences seen at 3~Myr compared to earlier times in the same model.
In both cases, the mechanism at work begins with carbon extraction from CO, which is initially placed in C$^+$.  C$^+$ is presumed to stay in the gas and not freeze-out due to the expectation that grains are negatively charged, which will result in recombination and gas phase ejection of the neutralized product\citep{Aikawa99b}.
 The C$^+$ ion is reactive and can gradually find its way into a variety of carbon bearing molecular ions.   The molecular ions will more rapidly dissociatively recombine with free electrons or on negatively charged grain surfaces to make a neutral molecule.   The carbon-bearing neutral molecule can deplete onto grain surfaces, provided the grain temperature is below its respective sublimation temperature.   However, if grain collision timescales are long the gas-phase chemistry has sufficient time to approach equilibrium which ultimately places the carbon back into CO.

Two points are notable here.  It is possible that CO chemical processing is only active at earlier stages prior to significant grain growth and as grains grow this mechanism becomes less effective.
 Along these lines the layer near the midplane where CO ice is frozen on grains is substantially reduced in the grain growth model, due to the decrease in the total grain surface area and a subsequent increase in the time-scale for freeze out.  The ice-dominated layer will be underestimated in this model, which does not include full time-dependent grain evolution, because grains will tend to trap ices as they grow.   In essence, whenever the smaller ice coated grains coagulate and grow to larger sizes it is expected that they maintain the pre-existing ice coatings.   An additional effect is the expected stratification in the grain size distribution with grains growing to larger sizes in the midplane and smaller grains lofted to greater heights due to coupling to gas motions\citep{dd04}.    
For instance, it is well known that small grains do exist on the surfaces of protoplanetary disks as scattering requires the presence of grains with smaller sizes, while the presence of larger grains is inferred via detection of centimeter-wave continuum excess emission\citep{wilner_twhya_cm}.   Thus the layers where CO chemical processing is active may vary in effectiveness depending on the available surface area of dust grains for gas phase depletion.   

\begin{figure}
\centering
    \includegraphics[width=0.9\textwidth, angle=0]{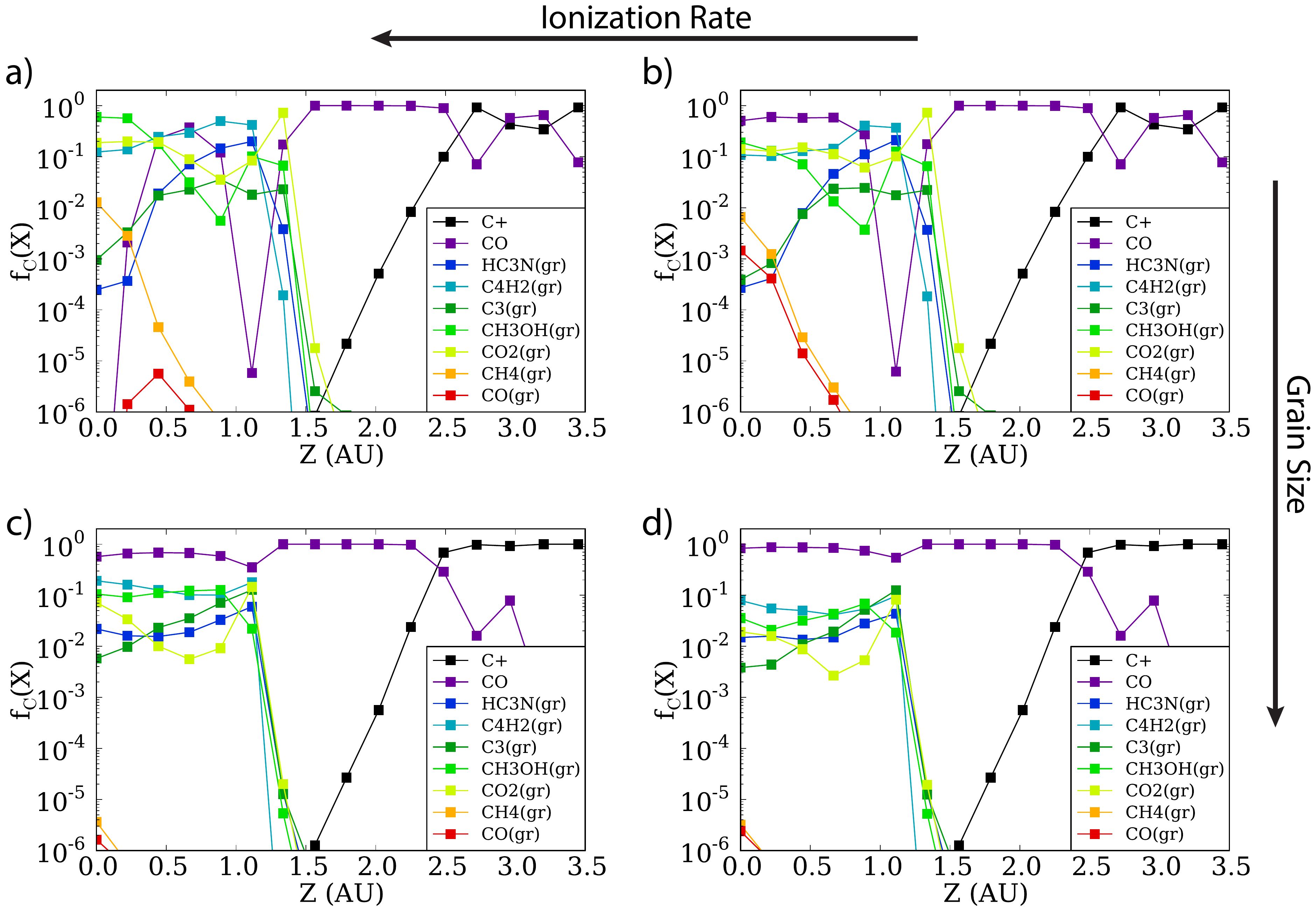}
     \caption{\footnotesize   Plot illustrating dominant volatile carbon carriers in either the gas phase or on the grain surface.  Plot is taken from the standard model at $\sim$1 Myr at a radius of 10 AU.  Abundances relative to the total amount of volatile carbon (i.e.  the initial CO abundance) are shown as a function of disk vertical height.
    All the colors are the same across the panels.  
  Panel (a) unattenuated cosmic ray radiation field and 0.1 $\mu$m grains; (b) model with wind-attenuated cosmic rays and 0.1 $\mu$m grains; (c) unattenuated cosmic ray radiation field and 10 $\mu$m grains; (d) model with wind attenuated cosmic rays and 10 $\mu$m grains.
 Arrows indicate trends.  }
  \label{fig:vert}
\end{figure}

\subsubsection{Possible Carbon Reservoirs in Disk Surface  }

In the preceding section we illustrated the diminishment of the CO gas phase abundance in gas exposed to ionization and a high total grain surface area.   The carbon extracted from CO via this mechanism is placed on grains in a variety of forms depending on the local physical environment.    In Fig.~\ref{fig:vert} we illustrate the effects of this process in a vertical cut of abundances at 10 AU, in the giant planet-forming zone taken at a timescale of 1 Myr.  In this plot we refer to abundances as percentages of volatile carbon.  Thus even an abundance of 10$^{-2}$ is high ($\sim 10^{-6}$, relative to H$_2$) by astronomical detection standards for gaseous molecules which are as low as 10$^{-11}$ in the dense ISM and perhaps somewhat higher in smaller disk systems.  Fig.~\ref{fig:vert}a presents the model with the disk exposed to an unattenuated cosmic ray radiation field with 0.1$\mu$m grains.  In this panel the vertical surface layer where this process is active is clearly seen as a drop in the CO abundance near Z $= 1.1$~AU.   Deeper in the disk, in the midplane, a similar depletion of gas phase CO is predicted.
At both points (Z $= 1.1$~AU and the midplane) there is a rise in the abundance of CO$_2$, CH$_3$OH, HC$_3$N, C$_3$, and C$_4$H$_2$ ice.    

   The production of these species proceeds through a strict gas phase process.  The destruction of CO via He$^+$ atoms creates C$^+$ ions.    It has been known for some time that chemical models, starting with carbon in C$^+$ within H$_2$ dominated gas, predict that, prior to the reformation of CO, the chemistry produces a host of complex molecular species\citep{glf82, lhh84}.  If the dust temperature is below the sublimation temperature for a given complex species then it will freeze onto the grain.   In this instance the disk densities are sufficiently high ($n > 10^{8}$~cm$^{-3}$) that the radiative association reaction between C$^+$ and H$_2$, creating CH$_2^+$, dominates along with a contribution of recombination on negatively charged grains.  One potential competitor to this process is the reaction between C$^+$ and H$_2$O, which provides an indirect channel to CO via HCO$^+$.    However, in these layers the dust grain temperature is well below that required for water sublimation ($> 100-150$~K, depending on pressure)\citep{fraser_h2obind}.     Instead a tenuous layer of water vapor is predicted to exist on deeper inside the exposed disk surface  induced by UV photodesorption, e.g. \citet{dchk05} and \citet{hkbm09}.   In our model shown in Fig.~\ref{fig:vert}a, C$^+$ destruction with H$_2$ is two orders of magnitude more effective than the reaction with photodesorbed water.   Here we have assumed a photodesorption yield of $10^{-3}$, motivated through the laboratory work by \citet{olvv10}.  A significantly higher yield would be required for this reaction to become important, which is inconsistent with existing water observations\citep{bergin10b, hoger11a}.  However, inside the water evaporation front this mechanism must therefore become inefficient.
     
  One beneficiary of this gas phase process is  CH$_3$OH which we use as an example of the end products.   When CO is destroyed,  gas phase formation of H$_2$CO leads to a production of H$_3$CO$^+$ via a reactionm with abundant HCO$^+$.  H$_3$CO$^+$ then reacts with CH$_4$ to form
CH$_3$OCH$_4^+$, which recombines with electrons in the gas or on grains to create gas phase CH$_3$OH, which freezes onto grains.
Similar sets of ion-molecule reactions lead to the other abundant ices.

In Fig.~\ref{fig:vert}b model results are presented assuming a reduced ionization rate by cosmic rays and a large surface area of grains.    At the surface the presence of strong X-ray ionization powers CO destruction and the creation of a similar suite of carbon-bearing molecules.    In the midplane the reduced level of ionization, which is now powered by scattered high energy X-rays, lead to less CO destruction and factors of a few less complex ice production.   However, we note that the process is still active and longer timescales will lead to activation.
Exploring models with reduced grain surface area (Figs.~\ref{fig:vert}c,d) we see that the effectiveness of this mechanism is significantly reduced, at least when viewed from the perspective of the CO abundance.     The reduction in the effectiveness of gas phase CO processing is due to longer freeze-out timescales allowing for significant CO re-formation to occur.
Regardless, processing still occurs at reduced levels providing astronomically significant amounts of a variety of complex species.  

\begin{figure}
\centering
    \includegraphics[width=0.9\textwidth, angle=0]{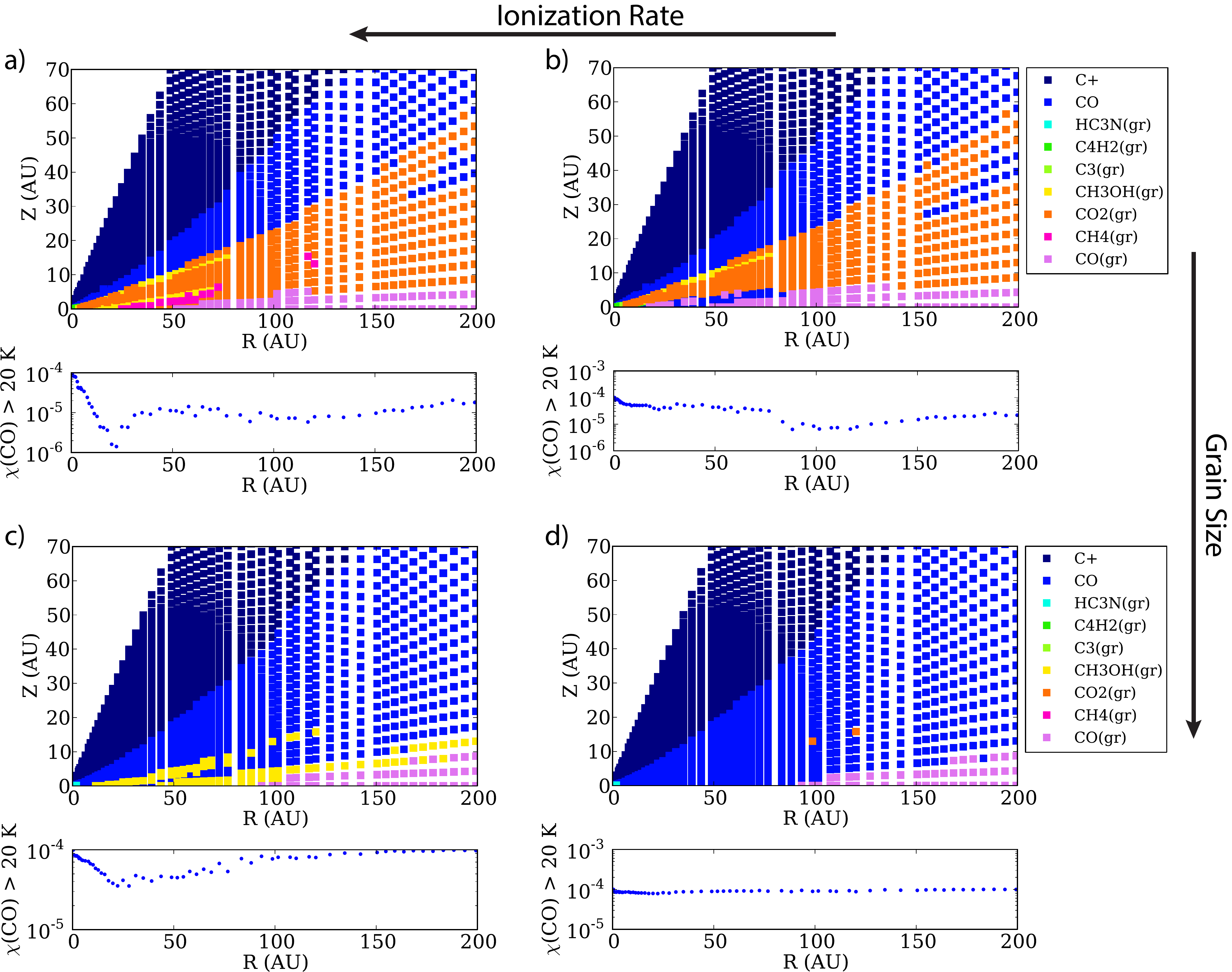}
      \caption{\footnotesize   Plot illustrating dominant volatile carbon carrier at each R,Z position in the gridded disk model.
  In this instance the dominant volatile carbon carrier contains over 50\% of the carbon that initially resided in CO (and
  sometimes is CO).
    All the colors are the same across the panels.  
 All panels present the standard model at 1~Myr.  Panel (a) unattenuated cosmic ray radiation field and 0.1 $\mu$m grains; (b) model with wind-attenuated cosmic rays and 0.1 $\mu$m grains; (c) unattenuated cosmic ray radiation field and 10 $\mu$m grains; (d) model with wind attenuated cosmic rays and 10 $\mu$m grains.
 Arrows indicate trends.  Underneath each panel the vertically averaged gas phase CO abundance above 20~K is shown as a function of radius for each respective model.}
   \label{fig:cres}
\end{figure}

In Fig.~\ref{fig:cres} the primary {\em volatile} carbon carriers are shown at 1~Myr for each of our four models as a function of disk radial and vertical position.   In the model with the highest ionization and largest surface grain surface area (Fig.~\ref{fig:cres}a), the general chemical structure on the surface of the disk for carbon species is seen for the outer disk:  C$^+$ $\rightarrow$ CO.  In the midplane CO is frozen on grains but above the midplane CO chemical processing leads primarily to the creation of CO$_2$ ice.    This can occur via two pathways.  In the gas a portion of the oxygen freed by CO destruction can react with H$_3^+$ ultimately leading (2/3 of the time) to the production of OH.  Existing gas phase CO will react with the hydroxyl producing CO$_2$, which can freeze onto grain surface.   An alternate route exists on grain surfaces in layers where CO is allowed to freeze onto grains in moderate amounts, at which point free OH on grain surfaces reacts with a small (80~K) energy barrier, directly making CO$_2$ ice.
  These are very localized on top of the regions where CO can be frozen onto grains and contributes at some level to the abundance of CH$_3$OH as well.

Similar results are seen for the model with a reduced cosmic ray ionization rate, but interstellar grain surface area (panel b).  In this model the active radionuclides provide a floor for the ionization rate still allowing for CO chemical processing throughout the disk.     One notable difference is seen in the midplane inside of the nominal CO evaporation front (near 70 AU) where CO evaporation and the presence of ionization leads to CO destruction and creation of CO$_2$ ice.   In the model with reduced ionization this effect is not observed, at least within the first 1~Myr.
The volatile CO abundance in the warm layers above the nominal freeze-out temperature of 20 K is substantially reduced in models with a high ionization rate.  If the ionization rate is reduced, the CO processing mechanism is more effective in the outer disk than the inner disk.  This is generally due to a lower equilibrium ion abundance at higher densities (i.e. more frequent recombination), thus leading to the ineffectiveness of midplane CO destruction in the inner 70 AU. 

In models with lower grain surface area (Fig.~\ref{fig:cres}c,d) the effectiveness of CO chemical processing is significantly reduced.   Beyond the midplane snowline the disk has the anticipated chemical structure for volatile carbon C$^+$ $\rightarrow$ CO $\rightarrow$ CO$_{gr}$.   
If the disk is directly exposed to interstellar cosmic rays there is some processing, which is more effective in the inner disk where freeze-out is relatively more efficient due to higher gas densities. This is most directly seen in the vertically averaged CO abundance in panel (c), where the CO abundance decreases to a minimum at 30 AU and then increases as the rising dust temperature reduces the number of available sinks on grain surface (i.e. complex molecules begin to evaporate). Thus the second part of the mechanism the sequestration of carbon on grain surfaces is hampered.  Finally in the model with low disk ionization and grain surface area little change is seen in the CO abundance.
 
In all models the dust temperature is what sets the specific carbon sink, as the chemistry tends to move towards the most abundant molecule the gas-phase chemistry can produce that will locally freeze-out.  Thus in the inner tens of AU a diversity of carbon carriers are predicted, e.g. HC$_3$N, C$_4$H$_2$, HC$_5$N, and C$_6$H$_2$. We stress that these predictions regarding the specific carbon carrier are highly uncertain as our model includes an incomplete surface chemistry.  Moreover there exist large uncertainties in the gas-grain interaction, as different binding energies for the various carbon carriers would change model predictions.   One intriguing facet is the general tendency of theoretical binding energies to be successively higher for larger molecules\citep{hhl92}, a facet that is broadly reproduced in lab experiments \citep{collings_lab, Oberg09a}.   This is consistent with the changing nature of the model predictions for the species that are the carbon-endpoints of this process. 
In summary, provided a source of ionization, across a range of dust temperatures, a variety of potential carbon-sinks for the chemistry can readily be created.  

\section{Implications for Carbon incorporation into Planetesimals and Planets}

In this contribution we have explored the potential for chemistry in the planet-forming disk to alter chemical abundances inherited from the interstellar medium.  In this case we refer to carbon monoxide, which forms in the gas long before stellar birth and is provided by collapse to the young protoplanetary disk.   The chemical processing mechanism requires grains to be warm ($>$ 20~K), partially ionized, and a total grain surface area per hydrogen above that of 1/1000 of the value in the interstellar medium (i.e. $n_g \sigma_g/n \simeq 2 \times 10^{-24}$ cm$^{-1}$).   If these conditions are met then the timescales for CO chemical destruction are short enough and there is sufficient ionization in the gas for ion-molecule reactions to slowly create a host of gas phase molecules with less volatility than CO that freeze onto grain surfaces.   For the most part this mechanism can extract carbon from CO within the disk surface layers, but can also be active in the disk midplane, inside the CO snowline.  Beyond the CO snowline any carbon locked as CO ice will remain unaltered, thus cometary CO is preserved.  A true disk is significantly more complex than explored in the static models presented here, as both the solids and the gaseous components can undergo systemic (advection) and semi-random (turbulent) motions.   In the case of the small grains with sizes $\sim 0.1\;\mu$m, they are coupled to the gas and via turbulence can migrate radially and vertically\citep{Ciesla12}, while intermediate size grains (cm to m-sizes) will undergo inward radial drift due to differential rotation speeds compared to the gaseous disk\citep{wc_ppiii}.    Thus there is substantial dynamical evolution which we have not explored.  This will influence the effectiveness of gas phase chemical processing of CO in both directions.     Finally there is the likelihood of the significant radial redistribution of materials as giant planets form, interact with the gaseous disk, leading to inward and outward movement depending on the timescales of gas dissipation.
Thus the ultimate composition of planetary materials may be comprised of material that originated at a variety of distances, e.g. \citet{walsh11}.

Observationally there is now evidence of below ISM CO abundance in the warm ($T > 20$~K) gas in one system, a result that we have shown is not due to differences in the self-shielding of HD and C$^{18}$O.  Similar effects need to be corroborated in other systems before concluding that this is a generic result of disk chemical evolution.   However, molecular ions, and stellar X-rays, are readily detected in disk systems confirming the presence of sources of ionization\citep{bergin_ppv, henning13}.  In the disk helium atoms will be in the vapor state; thus He$^+$ ions are present to react with CO on short timescales given the high densities.  The only question is how effective grains are at providing a sink.  In this light there are a number of additional chemical effects that are unexplored in our models that could lead to increased chemical complexity of the molecules species created from volatile carbon.
Beyond CH$_3$OH, we have not fully explored the potential for grain surface chemistry itself to lead to the creation of more complex organics with stronger bonds to the grain surface.  In warmer regions with T $\sim 20-30$~K grain surfaces are more conducive to a larger number of grain surface reaction pathways as heavier radicals become free to move and scan the surface\citep{gwh08}. 
Moreover, laboratory experiments of interstellar-analog molecular ices (e.g. H$_2$O, CH$_3$OH, CO) show that exposure to high levels of UV flux fosters the formation of a wide array of very complex organics similar to those found in meteorites\citep{mbernstein95, nuevo11}.   In addition \citet{Ciesla12} demonstrated that the dynamical motions of 1.0 $\mu$m grains leads to the likelihood that the surface ices are subject to repeated exposure to UV photons at levels similar to the laboratory experiments.  Thus the volatile carbon nominally locked in CO could readily be a source of the organic matter seen in meteoritic bodies.

 It is important to note that the question of carbon incorporated into terrestrial worlds is complex as it depends on the source terms (carbon-rich rocks or icy comets) and on whether the supplied carbon remains near the surface of the forming planet or is sequestered deeper into its core.   Our exploration of the fate of volatile carbon and its possible relation to the beginnings of organics in meteorites or comets is only one potential piece of this puzzle.  We have not discussed the fate of solid state or refractory carbon, such as amorphous carbon grains or PAHs which is another important aspect to consider\citep{lbn10, kress10, blo10}.    Looking forward, with the Atacama Large Millimeter Array in operation, there will be more data to compare to models to directly explore the fate of molecules formed in the dense interstellar medium.    Thus, the results presented here are the beginnings of our more direct exploration of the origins of material to be supplied to forming planets.
 


\section*{Acknowledgements}
This work was supported by funding from the National Science Foundation grant AST-1008800 and
AST-1344133 (INSPIRE).



\newcommand{\nat}{{ Nature }}
\newcommand{\aap}{{ Astron. \& Astrophys. }}
\newcommand{\aapr}{{ Astron. \& Astrophys. Rev.}}
\newcommand{\aj}{{ Astron.~J. }}
\newcommand{\apj}{{ Astrophys.~J. }}
\newcommand{\araa}{{ Ann. Rev. Astron. Astrophys. }}
\newcommand{\apjl}{{ Astrophys.~J.~Letters }}
\newcommand{\gca}{{ Geochim. Cosmochim. Acta}}
\newcommand{\ssr}{{ Space Science Rev.}}
\newcommand{\apjs}{{ Astrophys.~J.~Suppl. }}
\newcommand{\apss}{{ Astrophys.~Space~Sci. }}
\newcommand{\icarus}{{ Icarus }}
\newcommand{\mnras}{{ MNRAS }}
\newcommand{\pasp}{{ Pub. Astron. Soc. Pacific }}
\newcommand{\planss}{{ Plan. Space Sci. }}
\newcommand{\physrep}{{ Phys. Rep.}}
\newcommand{\bain}{{Bull.~Astron.~Inst.~Netherlands }}
\newcommand{\jgr}{{J. Geophys. Res.}}
\newcommand{\jcp}{{J. Chem. Phys.}}

\footnotesize{
\bibliography{ted} 

\providecommand*{\mcitethebibliography}{\thebibliography}
\csname @ifundefined\endcsname{endmcitethebibliography}
{\let\endmcitethebibliography\endthebibliography}{}
\begin{mcitethebibliography}{79}
\providecommand*{\natexlab}[1]{#1}
\providecommand*{\mciteSetBstSublistMode}[1]{}
\providecommand*{\mciteSetBstMaxWidthForm}[2]{}
\providecommand*{\mciteBstWouldAddEndPuncttrue}
  {\def\EndOfBibitem{\unskip.}}
\providecommand*{\mciteBstWouldAddEndPunctfalse}
  {\let\EndOfBibitem\relax}
\providecommand*{\mciteSetBstMidEndSepPunct}[3]{}
\providecommand*{\mciteSetBstSublistLabelBeginEnd}[3]{}
\providecommand*{\EndOfBibitem}{}
\mciteSetBstSublistMode{f}
\mciteSetBstMaxWidthForm{subitem}
{(\emph{\alph{mcitesubitemcount}})}
\mciteSetBstSublistLabelBeginEnd{\mcitemaxwidthsubitemform\space}
{\relax}{\relax}

\bibitem[{Morbidelli} \emph{et~al.}(2012){Morbidelli}, {Lunine}, {O'Brien},
  {Raymond}, and {Walsh}]{Morby12}
A.~{Morbidelli}, J.~I. {Lunine}, D.~P. {O'Brien}, S.~N. {Raymond} and K.~J.
  {Walsh}, \emph{Annual Review of Earth and Planetary Sciences}, 2012,
  \textbf{40}, 251--275\relax
\mciteBstWouldAddEndPuncttrue
\mciteSetBstMidEndSepPunct{\mcitedefaultmidpunct}
{\mcitedefaultendpunct}{\mcitedefaultseppunct}\relax
\EndOfBibitem
\bibitem[{van Dishoeck} \emph{et~al.}(2013){van Dishoeck}, {Bergin}, {Lis}, and
  {Lunine}]{evd_ppvi}
E.~{van Dishoeck}, E.~A. {Bergin}, D.~C. {Lis} and J.~I. {Lunine},
  \emph{Protostars and Planets VI}, 2013,  in press.\relax
\mciteBstWouldAddEndPunctfalse
\mciteSetBstMidEndSepPunct{\mcitedefaultmidpunct}
{}{\mcitedefaultseppunct}\relax
\EndOfBibitem
\bibitem[{Lee} \emph{et~al.}(2010){Lee}, {Bergin}, and {Nomura}]{lbn10}
J.-E. {Lee}, E.~A. {Bergin} and H.~{Nomura}, \emph{\apjl}, 2010, \textbf{710},
  L21--L25\relax
\mciteBstWouldAddEndPuncttrue
\mciteSetBstMidEndSepPunct{\mcitedefaultmidpunct}
{\mcitedefaultendpunct}{\mcitedefaultseppunct}\relax
\EndOfBibitem
\bibitem[{Pontoppidan} \emph{et~al.}(2013){Pontoppidan}, {Salyk}, {Bergin},
  {Brittain}, {Marty}, {Mousis}, and {\"{O}berg}]{pont_ppvi}
K.~{Pontoppidan}, C.~{Salyk}, E.~A. {Bergin}, S.~{Brittain}, B.~{Marty},
  O.~{Mousis} and K.~I. {\"{O}berg}, \emph{Protostars and Planets VI}, 2013,
  in press.\relax
\mciteBstWouldAddEndPunctfalse
\mciteSetBstMidEndSepPunct{\mcitedefaultmidpunct}
{}{\mcitedefaultseppunct}\relax
\EndOfBibitem
\bibitem[{Marty}(2012)]{Marty12}
B.~{Marty}, \emph{Earth and Planetary Science Letters}, 2012, \textbf{313},
  56--66\relax
\mciteBstWouldAddEndPuncttrue
\mciteSetBstMidEndSepPunct{\mcitedefaultmidpunct}
{\mcitedefaultendpunct}{\mcitedefaultseppunct}\relax
\EndOfBibitem
\bibitem[{Albarede} \emph{et~al.}(2013){Albarede}, {Ballhaus}, {Blichert-Toft},
  {Lee}, {Marty}, {Moynier}, and {Yin}]{Albarede13}
F.~{Albarede}, C.~{Ballhaus}, J.~{Blichert-Toft}, C.-T. {Lee}, B.~{Marty},
  F.~{Moynier} and Q.-Z. {Yin}, \emph{\icarus}, 2013, \textbf{222},
  44--52\relax
\mciteBstWouldAddEndPuncttrue
\mciteSetBstMidEndSepPunct{\mcitedefaultmidpunct}
{\mcitedefaultendpunct}{\mcitedefaultseppunct}\relax
\EndOfBibitem
\bibitem[{Favre} \emph{et~al.}(2013){Favre}, {Cleeves}, {Bergin}, {Qi}, and
  {Blake}]{favre13a}
C.~{Favre}, L.~I. {Cleeves}, E.~A. {Bergin}, C.~{Qi} and G.~A. {Blake},
  \emph{\apjl}, 2013, \textbf{776}, L38\relax
\mciteBstWouldAddEndPuncttrue
\mciteSetBstMidEndSepPunct{\mcitedefaultmidpunct}
{\mcitedefaultendpunct}{\mcitedefaultseppunct}\relax
\EndOfBibitem
\bibitem[{Aikawa} \emph{et~al.}(1998){Aikawa}, {Umebayashi}, {Nakano}, and
  {Miyama}]{Aikawa98}
Y.~{Aikawa}, T.~{Umebayashi}, T.~{Nakano} and S.~{Miyama}, \emph{Faraday
  Discussions}, 1998, \textbf{109}, 281\relax
\mciteBstWouldAddEndPuncttrue
\mciteSetBstMidEndSepPunct{\mcitedefaultmidpunct}
{\mcitedefaultendpunct}{\mcitedefaultseppunct}\relax
\EndOfBibitem
\bibitem[{Dasgupta} and {Hirschmann}(2010)]{dh10}
R.~{Dasgupta} and M.~M. {Hirschmann}, \emph{Earth and Planetary Science
  Letters}, 2010, \textbf{298}, 1--13\relax
\mciteBstWouldAddEndPuncttrue
\mciteSetBstMidEndSepPunct{\mcitedefaultmidpunct}
{\mcitedefaultendpunct}{\mcitedefaultseppunct}\relax
\EndOfBibitem
\bibitem[{Crockett} \emph{et~al.}(2014){Crockett}, {Bergin}, {Neill},
  {Favre},\emph{et~al.}]{Crockett14a}
N.~R. {Crockett}, E.~A. {Bergin}, J.~L. {Neill}, C.~{Favre} \emph{et~al.},
  \emph{\apj}, 2014,  submitted\relax
\mciteBstWouldAddEndPuncttrue
\mciteSetBstMidEndSepPunct{\mcitedefaultmidpunct}
{\mcitedefaultendpunct}{\mcitedefaultseppunct}\relax
\EndOfBibitem
\bibitem[{Adams}(2010)]{solarbirth}
F.~C. {Adams}, \emph{\araa}, 2010, \textbf{48}, 47--85\relax
\mciteBstWouldAddEndPuncttrue
\mciteSetBstMidEndSepPunct{\mcitedefaultmidpunct}
{\mcitedefaultendpunct}{\mcitedefaultseppunct}\relax
\EndOfBibitem
\bibitem[Nieva and Przybilla(2012)]{Nieva12}
M.~F. Nieva and N.~Przybilla, \emph{Astronomy and Astrophysics}, 2012,
  \textbf{539}, A143\relax
\mciteBstWouldAddEndPuncttrue
\mciteSetBstMidEndSepPunct{\mcitedefaultmidpunct}
{\mcitedefaultendpunct}{\mcitedefaultseppunct}\relax
\EndOfBibitem
\bibitem[{Plume} \emph{et~al.}(2012){Plume}, {Bergin}, {Phillips}, {Lis},
  {Wang}, {Crockett}, {Caux}, {Comito}, {Goldsmith}, and {Schilke}]{Plume12}
R.~{Plume}, E.~A. {Bergin}, T.~G. {Phillips}, D.~C. {Lis}, S.~{Wang}, N.~R.
  {Crockett}, E.~{Caux}, C.~{Comito}, P.~F. {Goldsmith} and P.~{Schilke},
  \emph{\apj}, 2012, \textbf{744}, 28\relax
\mciteBstWouldAddEndPuncttrue
\mciteSetBstMidEndSepPunct{\mcitedefaultmidpunct}
{\mcitedefaultendpunct}{\mcitedefaultseppunct}\relax
\EndOfBibitem
\bibitem[{Fomenkova}(1999)]{Fomenkova99}
M.~N. {Fomenkova}, \emph{\ssr}, 1999, \textbf{90}, 109--114\relax
\mciteBstWouldAddEndPuncttrue
\mciteSetBstMidEndSepPunct{\mcitedefaultmidpunct}
{\mcitedefaultendpunct}{\mcitedefaultseppunct}\relax
\EndOfBibitem
\bibitem[{Bergin} \emph{et~al.}(2013){Bergin}, {Cleeves}, {Gorti}, {Zhang},
  {Blake}, {Green}, {Andrews}, {Evans}, {Henning}, {{\"O}berg}, {Pontoppidan},
  {Qi}, {Salyk}, and {van Dishoeck}]{bergin_hd}
E.~A. {Bergin}, L.~I. {Cleeves}, U.~{Gorti}, K.~{Zhang}, G.~A. {Blake}, J.~D.
  {Green}, S.~M. {Andrews}, N.~J. {Evans}, II, T.~{Henning}, K.~{{\"O}berg},
  K.~{Pontoppidan}, C.~{Qi}, C.~{Salyk} and E.~F. {van Dishoeck}, \emph{\nat},
  2013, \textbf{493}, 644--646\relax
\mciteBstWouldAddEndPuncttrue
\mciteSetBstMidEndSepPunct{\mcitedefaultmidpunct}
{\mcitedefaultendpunct}{\mcitedefaultseppunct}\relax
\EndOfBibitem
\bibitem[{Williams} and {Cieza}(2011)]{williams_araa}
J.~P. {Williams} and L.~A. {Cieza}, \emph{\araa}, 2011, \textbf{49},
  67--117\relax
\mciteBstWouldAddEndPuncttrue
\mciteSetBstMidEndSepPunct{\mcitedefaultmidpunct}
{\mcitedefaultendpunct}{\mcitedefaultseppunct}\relax
\EndOfBibitem
\bibitem[{Mamajek}(2005)]{Mamajek05}
E.~E. {Mamajek}, \emph{\apj}, 2005, \textbf{634}, 1385--1394\relax
\mciteBstWouldAddEndPuncttrue
\mciteSetBstMidEndSepPunct{\mcitedefaultmidpunct}
{\mcitedefaultendpunct}{\mcitedefaultseppunct}\relax
\EndOfBibitem
\bibitem[{Linsky}(1998)]{linsky98}
J.~L. {Linsky}, \emph{\ssr}, 1998, \textbf{84}, 285--296\relax
\mciteBstWouldAddEndPuncttrue
\mciteSetBstMidEndSepPunct{\mcitedefaultmidpunct}
{\mcitedefaultendpunct}{\mcitedefaultseppunct}\relax
\EndOfBibitem
\bibitem[{Gorti} \emph{et~al.}(2011){Gorti}, {Hollenbach}, {Najita}, and
  {Pascucci}]{gorti11}
U.~{Gorti}, D.~{Hollenbach}, J.~{Najita} and I.~{Pascucci}, \emph{\apj}, 2011,
  \textbf{735}, 90\relax
\mciteBstWouldAddEndPuncttrue
\mciteSetBstMidEndSepPunct{\mcitedefaultmidpunct}
{\mcitedefaultendpunct}{\mcitedefaultseppunct}\relax
\EndOfBibitem
\bibitem[{Aikawa} \emph{et~al.}(2002){Aikawa}, {van Zadelhoff}, {van Dishoeck},
  and {Herbst}]{aikawa_vanz02}
Y.~{Aikawa}, G.~J. {van Zadelhoff}, E.~F. {van Dishoeck} and E.~{Herbst},
  \emph{\aap}, 2002, \textbf{386}, 622--632\relax
\mciteBstWouldAddEndPuncttrue
\mciteSetBstMidEndSepPunct{\mcitedefaultmidpunct}
{\mcitedefaultendpunct}{\mcitedefaultseppunct}\relax
\EndOfBibitem
\bibitem[{Williams} and {Best}(2013)]{wb13}
J.~P. {Williams} and W.~M.~J. {Best}, \emph{ApJ}, 2013,  submitted\relax
\mciteBstWouldAddEndPuncttrue
\mciteSetBstMidEndSepPunct{\mcitedefaultmidpunct}
{\mcitedefaultendpunct}{\mcitedefaultseppunct}\relax
\EndOfBibitem
\bibitem[{Wolcott-Green} and {Haiman}(2011)]{Wolcott-green11}
J.~{Wolcott-Green} and Z.~{Haiman}, \emph{\mnras}, 2011, \textbf{412},
  2603--2616\relax
\mciteBstWouldAddEndPuncttrue
\mciteSetBstMidEndSepPunct{\mcitedefaultmidpunct}
{\mcitedefaultendpunct}{\mcitedefaultseppunct}\relax
\EndOfBibitem
\bibitem[{Visser} \emph{et~al.}(2009){Visser}, {van Dishoeck}, and
  {Black}]{visser09}
R.~{Visser}, E.~F. {van Dishoeck} and J.~H. {Black}, \emph{\aap}, 2009,
  \textbf{503}, 323--343\relax
\mciteBstWouldAddEndPuncttrue
\mciteSetBstMidEndSepPunct{\mcitedefaultmidpunct}
{\mcitedefaultendpunct}{\mcitedefaultseppunct}\relax
\EndOfBibitem
\bibitem[{Hoff} \emph{et~al.}(1998){Hoff}, {Henning}, and {Pfau}]{hoff98}
W.~{Hoff}, T.~{Henning} and W.~{Pfau}, \emph{\aap}, 1998, \textbf{336},
  242--250\relax
\mciteBstWouldAddEndPuncttrue
\mciteSetBstMidEndSepPunct{\mcitedefaultmidpunct}
{\mcitedefaultendpunct}{\mcitedefaultseppunct}\relax
\EndOfBibitem
\bibitem[{Vacca} and {Sandell}(2011)]{vs11}
W.~D. {Vacca} and G.~{Sandell}, \emph{\apj}, 2011, \textbf{732}, 8\relax
\mciteBstWouldAddEndPuncttrue
\mciteSetBstMidEndSepPunct{\mcitedefaultmidpunct}
{\mcitedefaultendpunct}{\mcitedefaultseppunct}\relax
\EndOfBibitem
\bibitem[{van Dishoeck} \emph{et~al.}(1993){van Dishoeck}, {Blake}, {Draine},
  and {Lunine}]{evd_ppiii}
E.~F. {van Dishoeck}, G.~A. {Blake}, B.~T. {Draine} and J.~I. {Lunine},
  Protostars and Planets III, 1993, pp. 163--241\relax
\mciteBstWouldAddEndPuncttrue
\mciteSetBstMidEndSepPunct{\mcitedefaultmidpunct}
{\mcitedefaultendpunct}{\mcitedefaultseppunct}\relax
\EndOfBibitem
\bibitem[{Bergin} and {Tafalla}(2007)]{bt_araa}
E.~A. {Bergin} and M.~{Tafalla}, \emph{\araa}, 2007, \textbf{45},
  339--396\relax
\mciteBstWouldAddEndPuncttrue
\mciteSetBstMidEndSepPunct{\mcitedefaultmidpunct}
{\mcitedefaultendpunct}{\mcitedefaultseppunct}\relax
\EndOfBibitem
\bibitem[{Hollenbach} \emph{et~al.}(2009){Hollenbach}, {Kaufman}, {Bergin}, and
  {Melnick}]{hkbm09}
D.~{Hollenbach}, M.~J. {Kaufman}, E.~A. {Bergin} and G.~J. {Melnick},
  \emph{\apj}, 2009, \textbf{690}, 1497--1521\relax
\mciteBstWouldAddEndPuncttrue
\mciteSetBstMidEndSepPunct{\mcitedefaultmidpunct}
{\mcitedefaultendpunct}{\mcitedefaultseppunct}\relax
\EndOfBibitem
\bibitem[{Mumma} and {Charnley}(2011)]{mc11}
M.~J. {Mumma} and S.~B. {Charnley}, \emph{\araa}, 2011, \textbf{49},
  471--524\relax
\mciteBstWouldAddEndPuncttrue
\mciteSetBstMidEndSepPunct{\mcitedefaultmidpunct}
{\mcitedefaultendpunct}{\mcitedefaultseppunct}\relax
\EndOfBibitem
\bibitem[{Dutrey} \emph{et~al.}(1997){Dutrey}, {Guilloteau}, and
  {Guelin}]{dgg97}
A.~{Dutrey}, S.~{Guilloteau} and M.~{Guelin}, \emph{\aap}, 1997, \textbf{317},
  L55--L58\relax
\mciteBstWouldAddEndPuncttrue
\mciteSetBstMidEndSepPunct{\mcitedefaultmidpunct}
{\mcitedefaultendpunct}{\mcitedefaultseppunct}\relax
\EndOfBibitem
\bibitem[{Kastner} \emph{et~al.}(1997){Kastner}, {Zuckerman}, {Weintraub}, and
  {Forveille}]{kastner_twhya}
J.~H. {Kastner}, B.~{Zuckerman}, D.~A. {Weintraub} and T.~{Forveille},
  \emph{Science}, 1997, \textbf{277}, 67--71\relax
\mciteBstWouldAddEndPuncttrue
\mciteSetBstMidEndSepPunct{\mcitedefaultmidpunct}
{\mcitedefaultendpunct}{\mcitedefaultseppunct}\relax
\EndOfBibitem
\bibitem[{Hogerheijde} \emph{et~al.}(2011){Hogerheijde}, {Bergin},
  {Brinch},\emph{et~al.}]{hoger11a}
M.~R. {Hogerheijde}, E.~A. {Bergin}, C.~{Brinch} \emph{et~al.}, \emph{Science},
  2011, \textbf{334}, 338--340\relax
\mciteBstWouldAddEndPuncttrue
\mciteSetBstMidEndSepPunct{\mcitedefaultmidpunct}
{\mcitedefaultendpunct}{\mcitedefaultseppunct}\relax
\EndOfBibitem
\bibitem[{Pontoppidan} \emph{et~al.}(2005){Pontoppidan}, {Dullemond}, {van
  Dishoeck}, {Blake}, {Boogert}, {Evans}, {Kessler-Silacci}, and
  {Lahuis}]{pont05}
K.~M. {Pontoppidan}, C.~P. {Dullemond}, E.~F. {van Dishoeck}, G.~A. {Blake},
  A.~C.~A. {Boogert}, N.~J. {Evans}, II, J.~E. {Kessler-Silacci} and
  F.~{Lahuis}, \emph{\apj}, 2005, \textbf{622}, 463--481\relax
\mciteBstWouldAddEndPuncttrue
\mciteSetBstMidEndSepPunct{\mcitedefaultmidpunct}
{\mcitedefaultendpunct}{\mcitedefaultseppunct}\relax
\EndOfBibitem
\bibitem[{Honda} \emph{et~al.}(2009){Honda}, {Inoue}, {Fukagawa}, {Oka},
  {Nakamoto}, {Ishii}, {Terada}, {Takato}, {Kawakita}, {Okamoto}, {Shibai},
  {Tamura}, {Kudo}, and {Itoh}]{Honda09}
M.~{Honda}, A.~K. {Inoue}, M.~{Fukagawa}, A.~{Oka}, T.~{Nakamoto}, M.~{Ishii},
  H.~{Terada}, N.~{Takato}, H.~{Kawakita}, Y.~K. {Okamoto}, H.~{Shibai},
  M.~{Tamura}, T.~{Kudo} and Y.~{Itoh}, \emph{\apjl}, 2009, \textbf{690},
  L110--L113\relax
\mciteBstWouldAddEndPuncttrue
\mciteSetBstMidEndSepPunct{\mcitedefaultmidpunct}
{\mcitedefaultendpunct}{\mcitedefaultseppunct}\relax
\EndOfBibitem
\bibitem[{Kamp} and {Dullemond}(2004)]{Kamp04}
I.~{Kamp} and C.~P. {Dullemond}, \emph{\apj}, 2004, \textbf{615},
  991--999\relax
\mciteBstWouldAddEndPuncttrue
\mciteSetBstMidEndSepPunct{\mcitedefaultmidpunct}
{\mcitedefaultendpunct}{\mcitedefaultseppunct}\relax
\EndOfBibitem
\bibitem[{Nomura} \emph{et~al.}(2007){Nomura}, {Aikawa}, {Tsujimoto},
  {Nakagawa}, and {Millar}]{nomura07}
H.~{Nomura}, Y.~{Aikawa}, M.~{Tsujimoto}, Y.~{Nakagawa} and T.~J. {Millar},
  \emph{\apj}, 2007, \textbf{661}, 334--353\relax
\mciteBstWouldAddEndPuncttrue
\mciteSetBstMidEndSepPunct{\mcitedefaultmidpunct}
{\mcitedefaultendpunct}{\mcitedefaultseppunct}\relax
\EndOfBibitem
\bibitem[{Bergin} \emph{et~al.}(2007){Bergin}, {Aikawa}, {Blake}, and {van
  Dishoeck}]{bergin_ppv}
E.~A. {Bergin}, Y.~{Aikawa}, G.~A. {Blake} and E.~F. {van Dishoeck}, Protostars
  and Planets V, 2007, p. 751\relax
\mciteBstWouldAddEndPuncttrue
\mciteSetBstMidEndSepPunct{\mcitedefaultmidpunct}
{\mcitedefaultendpunct}{\mcitedefaultseppunct}\relax
\EndOfBibitem
\bibitem[{Hayashi}(1981)]{hayashi_mmsn}
C.~{Hayashi}, \emph{Progress of Theoretical Physics Supplement}, 1981,
  \textbf{70}, 35--53\relax
\mciteBstWouldAddEndPuncttrue
\mciteSetBstMidEndSepPunct{\mcitedefaultmidpunct}
{\mcitedefaultendpunct}{\mcitedefaultseppunct}\relax
\EndOfBibitem
\bibitem[{Collings} \emph{et~al.}(2003){Collings}, {Dever}, {Fraser}, and
  {McCoustra}]{collings_cobind}
M.~P. {Collings}, J.~W. {Dever}, H.~J. {Fraser} and M.~R.~S. {McCoustra},
  \emph{\apss}, 2003, \textbf{285}, 633--659\relax
\mciteBstWouldAddEndPuncttrue
\mciteSetBstMidEndSepPunct{\mcitedefaultmidpunct}
{\mcitedefaultendpunct}{\mcitedefaultseppunct}\relax
\EndOfBibitem
\bibitem[{Collings} \emph{et~al.}(2004){Collings}, {Anderson}, {Chen}, {Dever},
  {Viti}, {Williams}, and {McCoustra}]{collings_lab}
M.~P. {Collings}, M.~A. {Anderson}, R.~{Chen}, J.~W. {Dever}, S.~{Viti}, D.~A.
  {Williams} and M.~R.~S. {McCoustra}, \emph{\mnras}, 2004, \textbf{354},
  1133--1140\relax
\mciteBstWouldAddEndPuncttrue
\mciteSetBstMidEndSepPunct{\mcitedefaultmidpunct}
{\mcitedefaultendpunct}{\mcitedefaultseppunct}\relax
\EndOfBibitem
\bibitem[{Fraser} \emph{et~al.}(2001){Fraser}, {Collings}, {McCoustra}, and
  {Williams}]{fraser_h2obind}
H.~J. {Fraser}, M.~P. {Collings}, M.~R.~S. {McCoustra} and D.~A. {Williams},
  \emph{\mnras}, 2001, \textbf{327}, 1165--1172\relax
\mciteBstWouldAddEndPuncttrue
\mciteSetBstMidEndSepPunct{\mcitedefaultmidpunct}
{\mcitedefaultendpunct}{\mcitedefaultseppunct}\relax
\EndOfBibitem
\bibitem[{Aikawa} \emph{et~al.}(1997){Aikawa}, {Umebayashi}, {Nakano}, and
  {Miyama}]{aikawa97}
Y.~{Aikawa}, T.~{Umebayashi}, T.~{Nakano} and S.~M. {Miyama}, \emph{\apjl},
  1997, \textbf{486}, L51+\relax
\mciteBstWouldAddEndPuncttrue
\mciteSetBstMidEndSepPunct{\mcitedefaultmidpunct}
{\mcitedefaultendpunct}{\mcitedefaultseppunct}\relax
\EndOfBibitem
\bibitem[{Fogel} \emph{et~al.}(2011){Fogel}, {Bethell}, {Bergin}, {Calvet}, and
  {Semenov}]{fogel11}
J.~K.~J. {Fogel}, T.~J. {Bethell}, E.~A. {Bergin}, N.~{Calvet} and
  D.~{Semenov}, \emph{\apj}, 2011, \textbf{726}, 29\relax
\mciteBstWouldAddEndPuncttrue
\mciteSetBstMidEndSepPunct{\mcitedefaultmidpunct}
{\mcitedefaultendpunct}{\mcitedefaultseppunct}\relax
\EndOfBibitem
\bibitem[{Cleeves} \emph{et~al.}(2011){Cleeves}, {Bergin}, {Bethell}, {Calvet},
  {Fogel}, {Sauter}, and {Wolf}]{cleeves11}
L.~I. {Cleeves}, E.~A. {Bergin}, T.~J. {Bethell}, N.~{Calvet}, J.~K.~J.
  {Fogel}, J.~{Sauter} and S.~{Wolf}, \emph{\apjl}, 2011, \textbf{743},
  L2\relax
\mciteBstWouldAddEndPuncttrue
\mciteSetBstMidEndSepPunct{\mcitedefaultmidpunct}
{\mcitedefaultendpunct}{\mcitedefaultseppunct}\relax
\EndOfBibitem
\bibitem[{Herczeg} \emph{et~al.}(2002){Herczeg}, {Linsky}, {Valenti},
  {Johns-Krull}, and {Wood}]{herczeg_twhya1}
G.~J. {Herczeg}, J.~L. {Linsky}, J.~A. {Valenti}, C.~M. {Johns-Krull} and B.~E.
  {Wood}, \emph{\apj}, 2002, \textbf{572}, 310--325\relax
\mciteBstWouldAddEndPuncttrue
\mciteSetBstMidEndSepPunct{\mcitedefaultmidpunct}
{\mcitedefaultendpunct}{\mcitedefaultseppunct}\relax
\EndOfBibitem
\bibitem[{Preibisch} \emph{et~al.}(2005){Preibisch}, {Kim}, {Favata},
  {Feigelson}, {Flaccomio}, {Getman}, {Micela}, {Sciortino}, {Stassun},
  {Stelzer}, and {Zinnecker}]{Preibisch05}
T.~{Preibisch}, Y.-C. {Kim}, F.~{Favata}, E.~D. {Feigelson}, E.~{Flaccomio},
  K.~{Getman}, G.~{Micela}, S.~{Sciortino}, K.~{Stassun}, B.~{Stelzer} and
  H.~{Zinnecker}, \emph{\apjs}, 2005, \textbf{160}, 401--422\relax
\mciteBstWouldAddEndPuncttrue
\mciteSetBstMidEndSepPunct{\mcitedefaultmidpunct}
{\mcitedefaultendpunct}{\mcitedefaultseppunct}\relax
\EndOfBibitem
\bibitem[{Bethell} and {Bergin}(2011)]{bb11b}
T.~J. {Bethell} and E.~A. {Bergin}, \emph{\apj}, 2011, \textbf{740}, 7\relax
\mciteBstWouldAddEndPuncttrue
\mciteSetBstMidEndSepPunct{\mcitedefaultmidpunct}
{\mcitedefaultendpunct}{\mcitedefaultseppunct}\relax
\EndOfBibitem
\bibitem[{Bethell} and {Bergin}(2011)]{bb11a}
T.~J. {Bethell} and E.~A. {Bergin}, \emph{\apj}, 2011, \textbf{739}, 78\relax
\mciteBstWouldAddEndPuncttrue
\mciteSetBstMidEndSepPunct{\mcitedefaultmidpunct}
{\mcitedefaultendpunct}{\mcitedefaultseppunct}\relax
\EndOfBibitem
\bibitem[{Schindhelm} \emph{et~al.}(2012){Schindhelm}, {France}, {Herczeg},
  {Bergin}, {Yang}, {Brown}, {Brown}, {Linsky}, and {Valenti}]{Schindhelm12}
E.~{Schindhelm}, K.~{France}, G.~J. {Herczeg}, E.~{Bergin}, H.~{Yang},
  A.~{Brown}, J.~M. {Brown}, J.~L. {Linsky} and J.~{Valenti}, \emph{\apjl},
  2012, \textbf{756}, L23\relax
\mciteBstWouldAddEndPuncttrue
\mciteSetBstMidEndSepPunct{\mcitedefaultmidpunct}
{\mcitedefaultendpunct}{\mcitedefaultseppunct}\relax
\EndOfBibitem
\bibitem[{Cleeves} \emph{et~al.}(2013){Cleeves}, {Adams}, and
  {Bergin}]{cleeves13a}
L.~I. {Cleeves}, F.~C. {Adams} and E.~A. {Bergin}, \emph{\apj}, 2013,
  \textbf{772}, 5\relax
\mciteBstWouldAddEndPuncttrue
\mciteSetBstMidEndSepPunct{\mcitedefaultmidpunct}
{\mcitedefaultendpunct}{\mcitedefaultseppunct}\relax
\EndOfBibitem
\bibitem[{Smith} \emph{et~al.}(2004){Smith}, {Herbst}, and {Chang}]{osu04}
I.~W.~M. {Smith}, E.~{Herbst} and Q.~{Chang}, \emph{\mnras}, 2004,
  \textbf{350}, 323--330\relax
\mciteBstWouldAddEndPuncttrue
\mciteSetBstMidEndSepPunct{\mcitedefaultmidpunct}
{\mcitedefaultendpunct}{\mcitedefaultseppunct}\relax
\EndOfBibitem
\bibitem[{Hasegawa} \emph{et~al.}(1992){Hasegawa}, {Herbst}, and
  {Leung}]{hhl92}
T.~I. {Hasegawa}, E.~{Herbst} and C.~M. {Leung}, \emph{\apjs}, 1992,
  \textbf{82}, 167--195\relax
\mciteBstWouldAddEndPuncttrue
\mciteSetBstMidEndSepPunct{\mcitedefaultmidpunct}
{\mcitedefaultendpunct}{\mcitedefaultseppunct}\relax
\EndOfBibitem
\bibitem[{Charnley} and {Rodgers}(2009)]{cr09}
S.~B. {Charnley} and S.~B. {Rodgers}, Astronomical Society of the Pacific
  Conference Series, 2009, pp. 29--+\relax
\mciteBstWouldAddEndPuncttrue
\mciteSetBstMidEndSepPunct{\mcitedefaultmidpunct}
{\mcitedefaultendpunct}{\mcitedefaultseppunct}\relax
\EndOfBibitem
\bibitem[{Igea} and {Glassgold}(1999)]{ig99}
J.~{Igea} and A.~E. {Glassgold}, \emph{\apj}, 1999, \textbf{518},
  848--858\relax
\mciteBstWouldAddEndPuncttrue
\mciteSetBstMidEndSepPunct{\mcitedefaultmidpunct}
{\mcitedefaultendpunct}{\mcitedefaultseppunct}\relax
\EndOfBibitem
\bibitem[{Mathis} \emph{et~al.}(1977){Mathis}, {Rumpl}, and {Nordsieck}]{mrn}
J.~S. {Mathis}, W.~{Rumpl} and K.~H. {Nordsieck}, \emph{\apj}, 1977,
  \textbf{217}, 425--433\relax
\mciteBstWouldAddEndPuncttrue
\mciteSetBstMidEndSepPunct{\mcitedefaultmidpunct}
{\mcitedefaultendpunct}{\mcitedefaultseppunct}\relax
\EndOfBibitem
\bibitem[{Umebayashi} and {Nakano}(2009)]{umebayashi09}
T.~{Umebayashi} and T.~{Nakano}, \emph{\apj}, 2009, \textbf{690}, 69--81\relax
\mciteBstWouldAddEndPuncttrue
\mciteSetBstMidEndSepPunct{\mcitedefaultmidpunct}
{\mcitedefaultendpunct}{\mcitedefaultseppunct}\relax
\EndOfBibitem
\bibitem[{Cleeves} \emph{et~al.}(2013){Cleeves}, {Adams}, {Bergin}, and
  {Visser}]{cleeves13b}
L.~I. {Cleeves}, F.~C. {Adams}, E.~A. {Bergin} and R.~{Visser}, \emph{\apj},
  2013, \textbf{777}, 28\relax
\mciteBstWouldAddEndPuncttrue
\mciteSetBstMidEndSepPunct{\mcitedefaultmidpunct}
{\mcitedefaultendpunct}{\mcitedefaultseppunct}\relax
\EndOfBibitem
\bibitem[{Jura} \emph{et~al.}(2013){Jura}, {Xu}, and {Young}]{Jura13}
M.~{Jura}, S.~{Xu} and E.~D. {Young}, \emph{\apjl}, 2013, \textbf{775},
  L41\relax
\mciteBstWouldAddEndPuncttrue
\mciteSetBstMidEndSepPunct{\mcitedefaultmidpunct}
{\mcitedefaultendpunct}{\mcitedefaultseppunct}\relax
\EndOfBibitem
\bibitem[{Barlow}(1984)]{Barlow84}
S.~E. {Barlow}, \emph{PhD thesis}, UNIVERSITY OF COLORADO AT BOULDER.,
  1984\relax
\mciteBstWouldAddEndPuncttrue
\mciteSetBstMidEndSepPunct{\mcitedefaultmidpunct}
{\mcitedefaultendpunct}{\mcitedefaultseppunct}\relax
\EndOfBibitem
\bibitem[{Laudenslager} \emph{et~al.}(1974){Laudenslager}, {Huntress}, and
  {Bowers}]{lhb74}
J.~B. {Laudenslager}, W.~T. {Huntress}, Jr. and M.~T. {Bowers}, \emph{\jcp},
  1974, \textbf{61}, 4600--4617\relax
\mciteBstWouldAddEndPuncttrue
\mciteSetBstMidEndSepPunct{\mcitedefaultmidpunct}
{\mcitedefaultendpunct}{\mcitedefaultseppunct}\relax
\EndOfBibitem
\bibitem[{Natta} \emph{et~al.}(2007){Natta}, {Testi}, {Calvet}, {Henning},
  {Waters}, and {Wilner}]{natta_ppv}
A.~{Natta}, L.~{Testi}, N.~{Calvet}, T.~{Henning}, R.~{Waters} and D.~{Wilner},
  \emph{Protostars and Planets V}, 2007,  767--781\relax
\mciteBstWouldAddEndPuncttrue
\mciteSetBstMidEndSepPunct{\mcitedefaultmidpunct}
{\mcitedefaultendpunct}{\mcitedefaultseppunct}\relax
\EndOfBibitem
\bibitem[{Aikawa} \emph{et~al.}(1999){Aikawa}, {Herbst}, and
  {Dzegilenko}]{Aikawa99b}
Y.~{Aikawa}, E.~{Herbst} and F.~N. {Dzegilenko}, \emph{\apj}, 1999,
  \textbf{527}, 262--265\relax
\mciteBstWouldAddEndPuncttrue
\mciteSetBstMidEndSepPunct{\mcitedefaultmidpunct}
{\mcitedefaultendpunct}{\mcitedefaultseppunct}\relax
\EndOfBibitem
\bibitem[{Dullemond} and {Dominik}(2004)]{dd04}
C.~P. {Dullemond} and C.~{Dominik}, \emph{\aap}, 2004, \textbf{421},
  1075--1086\relax
\mciteBstWouldAddEndPuncttrue
\mciteSetBstMidEndSepPunct{\mcitedefaultmidpunct}
{\mcitedefaultendpunct}{\mcitedefaultseppunct}\relax
\EndOfBibitem
\bibitem[{Wilner} \emph{et~al.}(2000){Wilner}, {Ho}, {Kastner}, and
  {Rodr{\'{\i}}guez}]{wilner_twhya_cm}
D.~J. {Wilner}, P.~T.~P. {Ho}, J.~H. {Kastner} and L.~F. {Rodr{\'{\i}}guez},
  \emph{\apjl}, 2000, \textbf{534}, L101--L104\relax
\mciteBstWouldAddEndPuncttrue
\mciteSetBstMidEndSepPunct{\mcitedefaultmidpunct}
{\mcitedefaultendpunct}{\mcitedefaultseppunct}\relax
\EndOfBibitem
\bibitem[{Graedel} \emph{et~al.}(1982){Graedel}, {Langer}, and
  {Frerking}]{glf82}
T.~E. {Graedel}, W.~D. {Langer} and M.~A. {Frerking}, \emph{\apjs}, 1982,
  \textbf{48}, 321--368\relax
\mciteBstWouldAddEndPuncttrue
\mciteSetBstMidEndSepPunct{\mcitedefaultmidpunct}
{\mcitedefaultendpunct}{\mcitedefaultseppunct}\relax
\EndOfBibitem
\bibitem[{Leung} \emph{et~al.}(1984){Leung}, {Herbst}, and {Huebner}]{lhh84}
C.~M. {Leung}, E.~{Herbst} and W.~F. {Huebner}, \emph{\apjs}, 1984,
  \textbf{56}, 231--256\relax
\mciteBstWouldAddEndPuncttrue
\mciteSetBstMidEndSepPunct{\mcitedefaultmidpunct}
{\mcitedefaultendpunct}{\mcitedefaultseppunct}\relax
\EndOfBibitem
\bibitem[{Dominik} \emph{et~al.}(2005){Dominik}, {Ceccarelli}, {Hollenbach},
  and {Kaufman}]{dchk05}
C.~{Dominik}, C.~{Ceccarelli}, D.~{Hollenbach} and M.~{Kaufman}, \emph{\apjl},
  2005, \textbf{635}, L85--L88\relax
\mciteBstWouldAddEndPuncttrue
\mciteSetBstMidEndSepPunct{\mcitedefaultmidpunct}
{\mcitedefaultendpunct}{\mcitedefaultseppunct}\relax
\EndOfBibitem
\bibitem[{{\"O}berg} \emph{et~al.}(2009){{\"O}berg}, {Linnartz}, {Visser}, and
  {van Dishoeck}]{olvv10}
K.~I. {{\"O}berg}, H.~{Linnartz}, R.~{Visser} and E.~F. {van Dishoeck},
  \emph{\apj}, 2009, \textbf{693}, 1209--1218\relax
\mciteBstWouldAddEndPuncttrue
\mciteSetBstMidEndSepPunct{\mcitedefaultmidpunct}
{\mcitedefaultendpunct}{\mcitedefaultseppunct}\relax
\EndOfBibitem
\bibitem[{Bergin} \emph{et~al.}(2010){Bergin}, {Hogerheijde},
  {Brinch},\emph{et~al.}]{bergin10b}
E.~A. {Bergin}, M.~R. {Hogerheijde}, C.~{Brinch} \emph{et~al.}, \emph{\aap},
  2010, \textbf{521}, L33+\relax
\mciteBstWouldAddEndPuncttrue
\mciteSetBstMidEndSepPunct{\mcitedefaultmidpunct}
{\mcitedefaultendpunct}{\mcitedefaultseppunct}\relax
\EndOfBibitem
\bibitem[{{\"O}berg} \emph{et~al.}(2009){{\"O}berg}, {Garrod}, {van Dishoeck},
  and {Linnartz}]{Oberg09a}
K.~I. {{\"O}berg}, R.~T. {Garrod}, E.~F. {van Dishoeck} and H.~{Linnartz},
  \emph{\aap}, 2009, \textbf{504}, 891--913\relax
\mciteBstWouldAddEndPuncttrue
\mciteSetBstMidEndSepPunct{\mcitedefaultmidpunct}
{\mcitedefaultendpunct}{\mcitedefaultseppunct}\relax
\EndOfBibitem
\bibitem[{Ciesla} and {Sandford}(2012)]{Ciesla12}
F.~J. {Ciesla} and S.~A. {Sandford}, \emph{Science}, 2012, \textbf{336},
  452--\relax
\mciteBstWouldAddEndPuncttrue
\mciteSetBstMidEndSepPunct{\mcitedefaultmidpunct}
{\mcitedefaultendpunct}{\mcitedefaultseppunct}\relax
\EndOfBibitem
\bibitem[{Weidenschilling} and {Cuzzi}(1993)]{wc_ppiii}
S.~J. {Weidenschilling} and J.~N. {Cuzzi}, Protostars and Planets III, 1993,
  pp. 1031--1060\relax
\mciteBstWouldAddEndPuncttrue
\mciteSetBstMidEndSepPunct{\mcitedefaultmidpunct}
{\mcitedefaultendpunct}{\mcitedefaultseppunct}\relax
\EndOfBibitem
\bibitem[{Walsh} \emph{et~al.}(2011){Walsh}, {Morbidelli}, {Raymond},
  {O'Brien}, and {Mandell}]{walsh11}
K.~J. {Walsh}, A.~{Morbidelli}, S.~N. {Raymond}, D.~P. {O'Brien} and A.~M.
  {Mandell}, \emph{\nat}, 2011, \textbf{475}, 206--209\relax
\mciteBstWouldAddEndPuncttrue
\mciteSetBstMidEndSepPunct{\mcitedefaultmidpunct}
{\mcitedefaultendpunct}{\mcitedefaultseppunct}\relax
\EndOfBibitem
\bibitem[{Henning} and {Semenov}(2013)]{henning13}
T.~{Henning} and D.~{Semenov}, \emph{ArXiv e-prints}, 2013\relax
\mciteBstWouldAddEndPuncttrue
\mciteSetBstMidEndSepPunct{\mcitedefaultmidpunct}
{\mcitedefaultendpunct}{\mcitedefaultseppunct}\relax
\EndOfBibitem
\bibitem[{Garrod} \emph{et~al.}(2008){Garrod}, {Weaver}, and {Herbst}]{gwh08}
R.~T. {Garrod}, S.~L.~W. {Weaver} and E.~{Herbst}, \emph{\apj}, 2008,
  \textbf{682}, 283--302\relax
\mciteBstWouldAddEndPuncttrue
\mciteSetBstMidEndSepPunct{\mcitedefaultmidpunct}
{\mcitedefaultendpunct}{\mcitedefaultseppunct}\relax
\EndOfBibitem
\bibitem[{Bernstein} \emph{et~al.}(1995){Bernstein}, {Sandford}, {Allamandola},
  {Chang}, and {Scharberg}]{mbernstein95}
M.~P. {Bernstein}, S.~A. {Sandford}, L.~J. {Allamandola}, S.~{Chang} and M.~A.
  {Scharberg}, \emph{\apj}, 1995, \textbf{454}, 327--+\relax
\mciteBstWouldAddEndPuncttrue
\mciteSetBstMidEndSepPunct{\mcitedefaultmidpunct}
{\mcitedefaultendpunct}{\mcitedefaultseppunct}\relax
\EndOfBibitem
\bibitem[{Nuevo} \emph{et~al.}(2011){Nuevo}, {Milam}, {Sandford}, {De
  Gregorio}, {Cody}, and {Kilcoyne}]{nuevo11}
M.~{Nuevo}, S.~N. {Milam}, S.~A. {Sandford}, B.~T. {De Gregorio}, G.~D. {Cody}
  and A.~L.~D. {Kilcoyne}, \emph{Advances in Space Research}, 2011,
  \textbf{48}, 1126--1135\relax
\mciteBstWouldAddEndPuncttrue
\mciteSetBstMidEndSepPunct{\mcitedefaultmidpunct}
{\mcitedefaultendpunct}{\mcitedefaultseppunct}\relax
\EndOfBibitem
\bibitem[{Kress} \emph{et~al.}(2010){Kress}, {Tielens}, and
  {Frenklach}]{kress10}
M.~E. {Kress}, A.~G.~G.~M. {Tielens} and M.~{Frenklach}, \emph{Advances in
  Space Research}, 2010, \textbf{46}, 44--49\relax
\mciteBstWouldAddEndPuncttrue
\mciteSetBstMidEndSepPunct{\mcitedefaultmidpunct}
{\mcitedefaultendpunct}{\mcitedefaultseppunct}\relax
\EndOfBibitem
\bibitem[{Bond} \emph{et~al.}(2010){Bond}, {Lauretta}, and {O'Brien}]{blo10}
J.~C. {Bond}, D.~S. {Lauretta} and D.~P. {O'Brien}, \emph{\icarus}, 2010,
  \textbf{205}, 321--337\relax
\mciteBstWouldAddEndPuncttrue
\mciteSetBstMidEndSepPunct{\mcitedefaultmidpunct}
{\mcitedefaultendpunct}{\mcitedefaultseppunct}\relax
\EndOfBibitem
\end{mcitethebibliography}
\bibliographystyle{rsc}
}

\end{document}